\documentclass[preprint,aps]{revtex4}

\usepackage{graphicx}

\begin{document}

%%%%%%%%%%%%%%%%%%%%%%%%%%%%%%%%%%%%%%%%%%%%%%%%%%%%%%%%%%%%%%%%%%%%%%%%%%%%%%
%%%%%%%%%%%%%%%%%%%%%%%%%%%%%%%%%%%%%%%%%%%%%%%%%%%%%%%%%%%%%%%%%%%%%%%%%%%%%%
%%%%%%%%%%%%%%%%%%%%%%%%%%%%%%%%%%%%%%%%%%%%%%%%%%%%%%%%%%%%%%%%%%%%%%%%%%%%%%

\title{On the dynamics of chemical reactions of negative ions}

\author{Jochen Mikosch}

\affiliation{National Research Council of Canada, Steacie Institute for
  Molecular Sciences, 100 Sussex Drive, Ottawa, Ontario K1A OR6, Canada}

\author{Matthias Weidem{\"u}ller}

\affiliation{Physikalisches Institut, Universit{\"a}t Heidelberg,
Philosophenweg 12, 69120 Heidelberg, Germany}

\author{Roland Wester}
\email{roland.wester@physik.uni-freiburg.de}

\affiliation{Physikalisches Institut, Universit{\"a}t Freiburg,
Hermann-Herder-Stra{\ss}e 3, 79104 Freiburg, Germany}

\date{\today}

\begin{abstract}
  This review discusses the dynamics of negative ion reactions with neutral
  molecules in the gas phase. Most anion-molecule reactions proceed via a
  qualitatively different interaction potential than cationic or neutral
  reactions. It has been and still is the goal of many experiments to
  understand these reaction dynamics and the different reaction mechanisms
  they lead to. We will show how rate coefficients and cross sections for
  anion-molecule reactions are measured and interpreted to yield information
  on the underlying dynamics. We will also present more detailed approaches
  that study either the transient reaction complex or the energy- and
  angle-resolved scattering of negative ions with neutral molecules. With the
  help of these different techniques many aspects of anion-molecule reaction
  dynamics could be unravelled in the last years. However, we are still far
  from a complete understanding of the complex molecular interplay that is at
  work during a negative ion reaction.
\end{abstract}

\maketitle

\tableofcontents

%%%%%%%%%%%%%%%%%%%%%%%%%%%%%%%%%%%%%%%%%%%%%%%%%%%%%%%%%%%%%%%%%%%%%%%%%%%%%%
%%%%%%%%%%%%%%%%%%%%%%%%%%%%%%%%%%%%%%%%%%%%%%%%%%%%%%%%%%%%%%%%%%%%%%%%%%%%%%
%%%%%%%%%%%%%%%%%%%%%%%%%%%%%%%%%%%%%%%%%%%%%%%%%%%%%%%%%%%%%%%%%%%%%%%%%%%%%%

\section{Introduction}

% chemical reactions
Questions about the nature of chemical reactions, why and how they proceed and
how this can be used to form certain desired chemical products are already
very old. In fact they are older than most other of the current research
topics in atomic and molecular physics or physical chemistry. The efforts to
answer these questions have lead to numerous achievements over the centuries,
starting maybe with the re-discovery of the discreet atomic structure of
matter, leading to the invention of chemical catalysis, and including the
understanding of the quantum mechanical nature of the chemical bond. Several
technological advances have fertilised experimental research on the dynamics
of chemical reactions. Besides the development of versatile tunable laser
source, one notes supersonic single and crossed molecular beams
\cite{lee1986:nl}, multi-dimensional momentum imaging and coincidence
detection \cite{heck1995:annu,whitaker:book,sanov2008:irpc}, ultrafast
time-resolved spectroscopy \cite{stolow2004:cr}, and, most recently, the
preparation of cold and ultracold atoms and molecules
\cite{doyle2004:epd,krems2005}.

% reaction dynamics
Today the study of the reaction dynamics of molecules has advanced to precise
quantum-state resolved scattering experiments in quantitative agreement with
high-level scattering calculations -- at least for reactive complexes that
involve no more than four atoms. If more atoms contribute to a chemical
reaction the dimensionality of the scattering process, i.\ e.\ the number of
degrees of freedom, increases beyond what quantum scattering calculations can
do on current computers. Also experiments are challenged by the growing
complexity, because individual quantum states are increasingly difficult to
separate in the initial as well as the final state of a scattering
event. However, it is precisely this ``complexity limit'' in chemical dynamics
that is driving a lot of research, because one would like to understand the
details of reactions that are of relevance in organic chemistry, in living
cells, or in the Earth's atmosphere. In such many-atom reactions new phenomena
may occur that lead to different reaction mechanisms than in triatomic
reactions.

% ion-molecule reactions
Within the large field of reaction dynamics an important division follows the
electric charge of the collision partners that participate in the
reaction. Besides the neutral-neutral reaction processes, the second largest
class are ion-molecule reactions. As will be shown later, the much stronger
ion-neutral interaction than the neutral-neutral van-der-Waals interaction
leads to qualitatively different behaviour. Ion-molecule reactions have often
very high rate coefficients and are thus important as soon as ions are present
in an environment. These environments include discharge plasmas
\cite{armentrout2000:adv}, combustion processes, the Earth's ionosphere
\cite{smith1995:msr}, and many other stellar and interstellar
regions \cite{herbst1998:adv,petrie2007:msr}. Also in the condensed phase,
more specifically in liquid solutions, ion-molecule reactions are
important. Reactions of ions are consequently studied almost as long as
neutral-neutral reactions. They require, however, significantly different
experimental techniques.

% anion-molecule reactions
A large portion of the research on ion-molecule reactions is devoted to
positively charged ions. In dilute plasmas, ionisation by electron impact or
by radiation will directly lead to the formation of cations and they should
therefore play an important role there. Nevertheless, also negatively charged
ions are important constituents of gas phase environments, for example in the
lower ionosphere of the Earth. Here the negative charge density is mostly
represented by atomic and molecular anions (such as O$^-$, O$_2^-$, OH$^-$,
HCO$_3^-$, NO$_2^-$ and their clusters with water molecules) and not by free
electrons \cite{smith1995:msr}. Similarly negative ions are omnipresent in
liquid solutions and have a strong impact on reactions there. Furthermore,
negative ions are precursors for solvated electrons in liquids, which are
responsible for strong optical absorption and have been the subject of many
studies.

% negative ions in the environment
In recent years negative ions have also been detected in other planetary
atmospheres in our solar system, such as the atmosphere of Saturn's satellite
titan, and they have been found in the interstellar medium, in the dark
molecular cloud TMC-1 and in a circumstellar shell around the carbon rich star
IRC+10216 \cite{mccarthy2006:apj,bruenken2007:apj}. The understanding of the
complex chemistry that leads to anions in these areas is still
fragmentary. The questions here are not only related to the formation of the
anions, but also if they can add new formation pathways to the complex neutral
molecules found in the interstellar medium, such as long unsaturated carbon
chains or possibly polyaromatic hydrocarbon molecules.

% topic of this review
The study of the reaction dynamics of anion-molecule reactions are the topic
of this review. Comparisons with cation-molecule reactions will be made if
suitable. Chemical reactions of negative ions have been studied extensively
for many decades. Several excellent reviews have appeared a few years ago on
this topic \cite{depuy2000:ijm,gronert2001:cr,ervin2001:cr}.  Here, we will
concentrate specifically on the reaction dynamics in contrast to the mere
kinetics of anion-molecule reactions. ``Molecular reaction dynamics is the
study of elementary processes [of molecular collisions] and the means of
probing them, understanding them and controlling them.''
\cite{levine2005}. In this sense we focus on processes where isolated
molecules collide in a well-controlled reactive event with a negative ion. We
give an overview over all the major experimental techniques that are used to
study anion-molecule reaction dynamics. In particular, we include the recent
experimental developments, on the one hand cryogenic ion traps and on the
other hand ion-molecule crossed-beam imaging, and show how they provide new
insights.

% separation from spectroscopic studies
Negative ions are also subject to spectroscopic studies in the infrared and
optical range, using bound-bound absorption and bound-free photoelectron
spectroscopy. This reveals the stability, i.\ e.\ the electron affinity of the
neutral, as well as structural information about the anion. Since negative
ions are still today challenging for quantum chemical calculations, due to the
spatial extension of the wavefunction of the excess electron and its
interaction with the atomic core, spectroscopy and detachment studies can
provide benchmark data
\cite{hammer2004:sci,sanov2008:irpc,neumark2008:jpc,wang2009:annu,hlavenka2009:jcp,gerardi2010:jpc}. Photoelectron
spectroscopy of negative ions can also reveal the energy level structure of
the neutral system that is reached after photodetaching the excess
electron. If the neutral system is not a bound molecule, but a reactive
complex, such as for the FH$_2^-$ anion, one can even study neutral reactions
near the transition state between reactants and products
\cite{neumark2005:pccp}. For more information on the spectroscopy of negative
ions see Refs.\ \cite{pegg2004:rpp,simons2008:jpc,wester2010:book}.

% structure of this article
This article will start with the discussion of a few general considerations of
anion-molecule collisions and reactions in the next section. Then it will be
discussed how measurements of integral cross sections or total rate
coefficients, respectively, are performed. They are carried out either at a
given collision energy or a given sample temperature and can already be used
to infer information on the underlying reaction dynamics. In the next section,
experimental approaches are discussed that allow the investigation of the
transient reaction complex. Then in section \ref{crossed-beams:sect}
experiments with crossed beams are presented, which allow the determination of
energy- and angle-differential cross sections, under optimum conditions even
quantum-state resolved. From these measurements a wealth of information about
the collision and reaction dynamics can be extracted. In the final section
some of the future directions for the study of anion-molecule reaction
dynamics are highlighted.

%%%%%%%%%%%%%%%%%%%%%%%%%%%%%%%%%%%%%%%%%%%%%%%%%%%%%%%%%%%%%%%%%%%%%%%%%%%%%%
%%%%%%%%%%%%%%%%%%%%%%%%%%%%%%%%%%%%%%%%%%%%%%%%%%%%%%%%%%%%%%%%%%%%%%%%%%%%%%
%%%%%%%%%%%%%%%%%%%%%%%%%%%%%%%%%%%%%%%%%%%%%%%%%%%%%%%%%%%%%%%%%%%%%%%%%%%%%%

\section{\label{interaction:sect} Anion-molecule interactions}

The understanding of the dynamics of an anion-molecule reaction starts with
the description of the interaction potential, usually in the Born-Oppenheimer
approximation. The long-range and short-range properties of the interaction
potential are discussed in the next sections. The proper theoretical treatment
of the dynamics of the colliding molecules and ions on the interaction
potential is a quantum scattering calculation
\cite{clary1998:sci,althorpe2003:annu,schmatz2004:cpc}, which is usually,
however, limited to a reduced set of dimensions by the computational
effort. Alternatively, the atomic motion is treated with classical dynamics
\cite{hase1994:sci,hase1998:adv_gas_chem}, which is generally a good
assumption because of the small de-Broglie wavelength of the atoms. In this
case all dimensions can be included, but such effects as the zero-point
motion, tunnelling through potential barriers or quantum scattering resonances
can usually not be accounted for.

%%%%%%%%%%%%%%%%%%%%%%%%%%%%%%%%%%%%%%%%%%%%%%%%%%%%%%%%%%%%%%%%%%%%%%%%%%%%%%

\subsection{Long-range and short range potentials}

% long-range electrostatic potential
At large distances the force between an ion and a neutral molecule is
determined by the electrostatic interaction potential of the ion with the
induced, and possibly also permanent, electric dipole moment of the molecule.
\begin{equation}
V(R) = - \frac{q^2 \alpha}{2 R^4} - \frac{q \mu}{R^2} \cos\theta
\label{long-range-pot:eq}
\end{equation}
$\alpha$ is the orientationally averaged polarisability and $\mu$ the
permanent dipole moment of the neutral molecule. $q$ is the charge of the
colliding ion. In certain cases also the ion-quadrupole interaction and other
higher order multipole terms need to be added.

% permanent dipole moment
If the molecule carries a permanent dipole moment its interaction with the
charge, which scales as $R^{-2}$ typically dominates over the induced-dipole
interaction, which scales as $R^{-4}$. However, at large relative separation
the free rotation of the molecule leads to a vanishing time-averaged permanent
dipole moment. At the largest inter-atomic distances the ion-induced dipole
interaction is therefore most important, which is always attractive. Once the
interaction potential then becomes comparable to the rotational energy
``locking'' of the permanent molecular dipole to the incoming ion may occur
and will thereby enhance the attractive interaction.

%%% Figure
\begin{figure}[p]
\center
\includegraphics[width=\columnwidth]{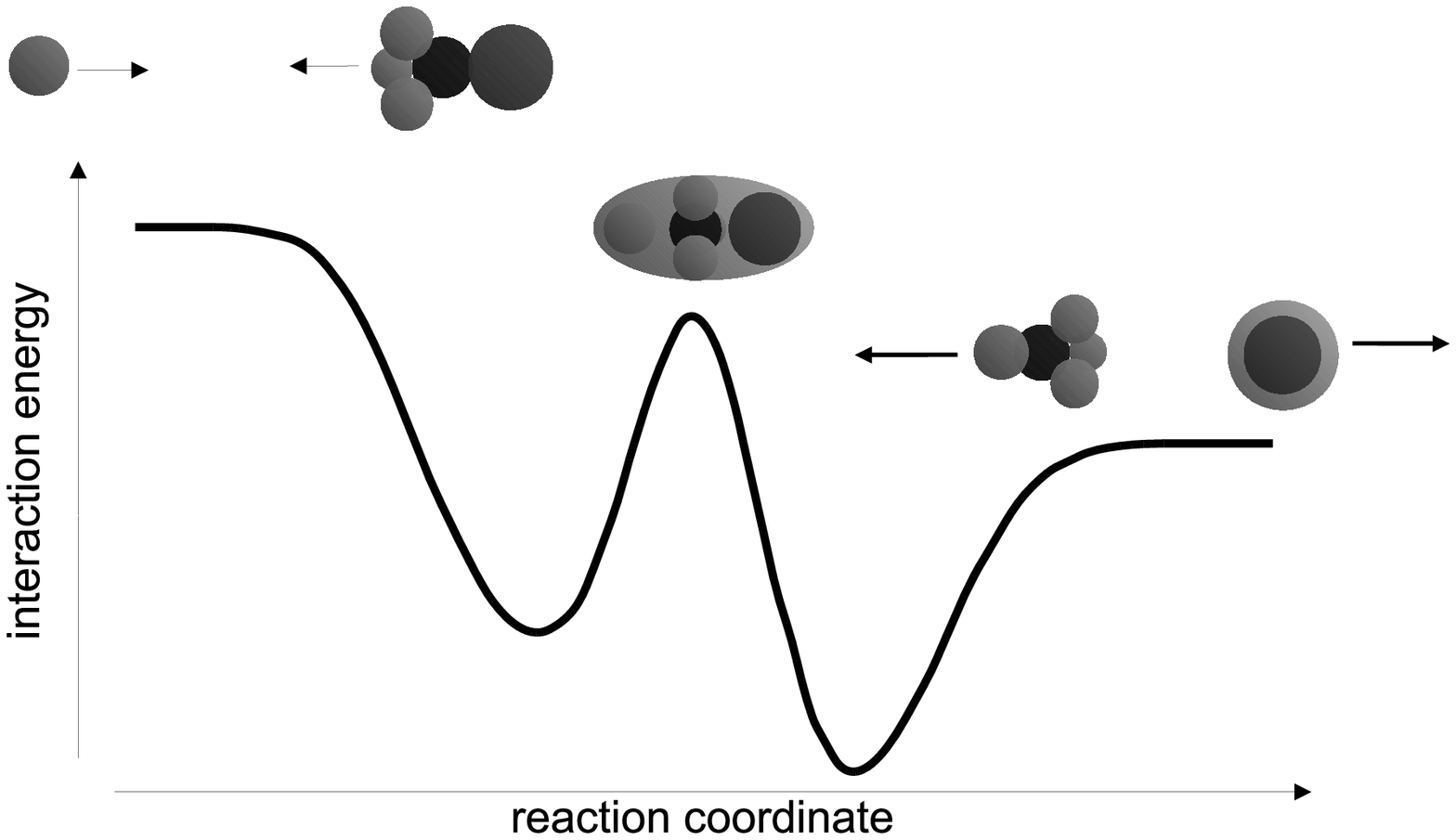}
\caption{Schematic view of the Born-Oppenheimer potential of an anion-molecule
  nucleophilic substitution reaction along the reaction path. This path is
  defined as the lowest-energy trajectory across the Born-Oppenheimer
  hypersurface that connects the reactants with the products. The presented
  potential exhibits two minima and an intermediate reaction barrier between
  them, which is a typical configuration for many anion-neutral
  reactions. Often the barrier is found below the energy of one of the
  asymptotes.}
\label{reaction-pot:fig}
\end{figure}

% short range: general statements for anions
At short distances the electronic wavefunctions of the ion and the molecule
interact. Determining the short range interaction potential therefore requires
solving the Schr{\"o}dinger equation for the collision system. This is often
particularly difficult for negative ions, because the excess electron is so
delocalized that a large basis set is required in the numerical solution of
the Schr{\"o}dinger equation. However, it works as an advantage that many
negative ions of interest are ``closed-shell'' systems with a spin singlet
configuration of the electronic wavefunction, because these are the most
stable negative ions with respect to electron detachment. As a consequence,
only a single Born-Oppenheimer potential hypersurface governs the reactions of
such negative ions with closed-shell neutral molecules.

% short range repulsive interaction, double-well potential surface
For many negative ion-neutral collision systems a repulsive potential is
obtained at very short range. This may be attributed to the electron-electron
repulsion by electrostatic forces and the Pauli exclusion principle. Together
with the strong attractive interaction this leads to potential energy
landscapes with two characteristic minima along the reaction coordinate, as
shown in Fig.\ \ref{reaction-pot:fig}. The path along the reaction coordinate
is defined as the path of minimum potential energy that connects the reactants
with the products. The potential energy minima typically lie several hundred
meV below the asymptotic energies. Consequently, substantial short-range
energy barriers may have negative transition state energies, as shown in Fig.\
\ref{reaction-pot:fig}. But even if the transition state energy is negative,
it can have a profound influence on the reactivity as we will see in the next
paragraph. This is in contrasts to most reactions of cations, because there is
usually no or at least no substantial short-range barrier.

%%%%%%%%%%%%%%%%%%%%%%%%%%%%%%%%%%%%%%%%%%%%%%%%%%%%%%%%%%%%%%%%%%%%%%%%%%%%%%

\subsection{Capture model}

% reaction rate coefficient
The kinetics of a chemical reaction, which describes the rate at which
products are formed as a function of time, is usually characterised by the
reaction rate coefficient. For a bimolecular reaction that occurs in a
collision of two species the second order rate equation
\begin{equation}
\frac{d\,n_{\rm product}}{dt} = k\,n_{\rm reactant 1}\,n_{\rm reactant 2}
\end{equation}
describes the time-dependent increase of the product particle density $n_{\rm
  product}$ for given reactant densities $n_{\rm reactant 1,2}$. This equation
is typically applied to thermal ensembles. The second order rate coefficient
$k$ is then dependent on the absolute temperature $T$. It is related to the
scattering cross section $\sigma(v_{\rm rel})$ for the individual collision
events, which depends on the relative velocity, by the thermal average
\begin{equation}
k 
= 
\langle \sigma(v_{\rm rel}) v_{\rm rel} \rangle_T
=
\int_0^\infty \sigma(v_{\rm rel}) p_T(v_{\rm rel}) v_{\rm rel} dv_{\rm rel}.
\end{equation}
$p_T(v_{\rm rel})$ is the thermal probability distribution for the relative
velocities at the absolute temperature $T$.

% langevin rate
The standard model to estimate collision rates of ions with neutral atoms or
molecules is the Langevin or capture model \cite{levine2005}. The assumption
of this model is that a collision occurs with 100\% probability if the two
collision partners come closer to each other than a critical distance. Only
the longest-range attractive interaction, the ion-induced dipole potential
(first term in Eq.\ \ref{long-range-pot:eq}), and the repulsive centrifugal
potential are taken into account. This critical distance is then given by the
location of the maximum of the centrifugal barrier. The largest impact
parameter for a scattering trajectory that reaches this critical distance and
can surmount the centrifugal barrier determines the scattering cross
section. For larger impact parameters the ion-neutral interaction is
neglected. With an additional thermal averaging, this assumption yields the
Langevin rate coefficient (in SI units)
\begin{equation} 
  k = \frac{|q|}{2 \epsilon_0} \sqrt{\frac{\alpha}{m_{\rm r}}},
  \label{langevin:eq}
\end{equation}
which turns out to be temperature independent. $\alpha$ is again the
orientation-averaged polarisability of the neutral molecule while $m_{\rm r}$
denotes the reduced mass of the two-body system ($q$ is the ion charge and
$\epsilon_0$ the electric constant). Typical Langevin rate coefficients range
between $5\times10^{−10}$\,cm$^3$/s and $5\times10^{−9}$\,cm$^3$/s.

% ion-polar molecule capture theory
A correction to the Langevin rate coefficient has to be introduced if the
neutral target carries a permanent dipole moment $\mu$. Then the second term
in Eq.\ (\ref{long-range-pot:eq}) needs to be included. The ``average dipole
orientation'' approach \cite{su1975:ijm} introduced an effective ``dipole
locking constant'' from which an improved collision rate coefficient was
derived. In a more accurate approach, the rate coefficient for ion-polar
molecule collisions is calculated from a series of classical trajectories that
are calculated to numerical precision for the exact interaction potential. The
parameterisation of these trajectories leads to a pre-factor $K(T)>1$ to the
Langevin rate constant Eq.\ \ref{langevin:eq}, which is a function of $\mu$
and $\alpha$ \cite{su1982:jcp,su1988:jcp,lim1994:qcpe}. As an effect the
ion-polar molecule capture rate coefficient is typically a factor of two to
four larger than the prediction for the Langevin rate coefficient.

%%%%%%%%%%%%%%%%%%%%%%%%%%%%%%%%%%%%%%%%%%%%%%%%%%%%%%%%%%%%%%%%%%%%%%%%%%%%%%
%%%%%%%%%%%%%%%%%%%%%%%%%%%%%%%%%%%%%%%%%%%%%%%%%%%%%%%%%%%%%%%%%%%%%%%%%%%%%%
%%%%%%%%%%%%%%%%%%%%%%%%%%%%%%%%%%%%%%%%%%%%%%%%%%%%%%%%%%%%%%%%%%%%%%%%%%%%%%

\section{Dynamics inferred from kinetics}

The Langevin rate constant has proven to be extremely valuable as a guideline
and often describes fairly accurately measured reaction rates for cations with
nonpolar neutral molecules \cite{ferguson1975:annu}. When the observed rate
coefficient for an ion-molecule reaction is smaller than the capture rate
coefficient, which occurs when the reaction probability after crossing the
centrifugal barrier is smaller than 100\%, temperature- or energy-dependent
rate coefficient measurements can be used to infer information on the reaction
dynamics at short internuclear separation.

In the next sections this approach will be illustrated with several studies
from recent years. Most major experimental techniques that are used for these
studies, drift and flow tubes, guided ion beams, free jet expansions and
low-temperature ion traps will be discussed. Ion cyclotron resonance
experiments and high-pressure mass spectrometry (see e.g.\
\cite{gronert2001:cr}), however, will not be covered in this article.

%%%%%%%%%%%%%%%%%%%%%%%%%%%%%%%%%%%%%%%%%%%%%%%%%%%%%%%%%%%%%%%%%%%%%%%%%%%%%%

\subsection{Reaction rate coefficients from drift tubes}

% flow and drift tube experiments
Flow and drift tubes have been the classic workhorse for the acquisition of
ion-molecule reaction rate coefficients. They have produced a wealth of
atmospherically and astrophysically relevant data over the almost 50 years of
their existence
\cite{ferguson1969:cjc,fehsenfeld1974:jcp,depuy1981:acr,lindinger1998:ijms,viggiano2001:agc}. Today
they are found in many different variants e.g.\
\cite{adams1976:ijms,vandoren1987:ijms,arnold2000:jpca,korolov2008:cpp}. These
instruments usually operate at room temperature, but may also be heated up to
more than 1000\,K or cooled down to near liquid nitrogen temperatures
(77\,K). Ions are created typically by electron impact, eventually chemically
transformed and mass selected, and then injected into a constant flow of
buffer gas of several m/s velocity. The pressure of this buffer gas, which is
usually helium, is of the order of a few millibar. Consequently, the mean free
path of the ions is smaller than the dimension of the tube and the flowing
buffer gas transports the thermalised ions downstream through the flow tube. A
distance away from the ion source, neutral reactant gas is injected. Due to
reactions of the ions with the neutral gas their density decreases with flow
time, which now corresponds to the flow distance. At the same time, product
ions are generated. By sampling ion yield and composition with a quadrupole
mass spectrometer as a function of flow distance absolute reaction rate
coefficients can be extracted.  Their temperature dependence may be obtained
by heating or cooling the buffer gas inlet and flow section. An electric field
gradient may be applied along the flow section turning the flow tube into a
drift tube. This allows studies at elevated kinetic energy rather than
temperature.

% anion-molecule reaction rates
The rate coefficient gives a first hint at the dynamics of an ion-molecule
reaction. Barrier-less proton transfer reactions usually exhibit large rate
coefficients in accordance with the capture model \cite{boehme2000:ijm}. In
contrast many anion-molecule reactions feature orders of magnitude lower rate
coefficients, pointing at dynamical bottlenecks and intermediate barriers on
the potential energy hypersurface. Chemically versatile flow and drift tubes
are well suited to study the effects of chemical substitution on the rate
coefficient \cite{depuy1990:jacs}. Different reaction mechanisms leading to
the same (ionic) products may be distinguished by studying kinetic isotope
effects \cite{eyet2008:jasms,villano2009:jacs}. More direct insight into the
dynamics can be obtained from a temperature dependent flow tube measurement of
the rate coefficient \cite{seeley1997:jacs,kerkines2010:jcp}. Moreover, in
drift tubes internal and translational degrees of freedom of the reaction
partners are decoupled and can be separately controlled. This allows to
investigate how the reaction rate changes if the same amount of energy is
provided in different forms, challenging a statistical model of the reaction
dynamics \cite{viggiano1992:jacs}.

% example 1: statistical redistribution in SN2
Probably the most important type of anion-molecule reaction that has been
studied up to now is the nucleophilic substitution (S$_{\rm N}$2) reaction
\cite{laehrdahl2002:ijms,uggerud2006:jpo}:
\begin{equation}
{\rm X}^- + {\rm CH}_3{\rm\,Y} \rightarrow {\rm Y}^- + {\rm CH}_3{\rm\,X},
\label{sn2:reaction}
\end{equation}
where X and Y can anything from simple halogen atoms to large macro-molecular
systems. The assumption of statistical energy redistribution was a matter of
fierce debate in the 1990s in the framework of S$_{\rm N}$2 reactions. A
variable temperature flow tube study had shown that the Cl$^-$ + CH$_3$Br
reaction rate is strongly dependent on the relative translational energy,
while being at the same time insensitive to the internal temperature of the
reactants \cite{viggiano1992:jacs}. Such behaviour is in clear contradiction
with the statistical assumption of rapid randomisation of all available energy
in the [Cl $\cdots$ CH$_3$Br]$^-$ entrance channel complex. This observation
prompted a large number of theoretical studies, which uncovered a dynamical
bottleneck for energy transfer between internal modes of CH$_3$Br and the
intermolecular low-frequency modes \cite{laehrdahl2002:ijms}.

% example 2: BrO- and ClO- SN2/E2
In a recent application of an ion flow tube instrument, the competition
between nucleophilic substitution (S$_{\rm N}$2) and base-induced elimination
(E2) was investigated via chemical substitution and the deuterium kinetic
isotope effect \cite{villano2006:jacs,villano2009:jacs}. Villano et
al. studied the reaction of BrO$^-$ and ClO$^-$ with methyl chloride
(CH$_3$Cl) and its partially and fully methylated form CXYZCl (where X,Y,Z can
be either H or CH$_3$). S$_{\rm N}$2 and E2 mechanisms lead to the same ionic
reaction product, a challenge for all experiments, which rely on charged
particle detection. The authors used that deuteration of the neutral reactant
changes the rate of the reaction. The deuterium kinetic isotope effect is
defined as the ratio of perprotio to perdeuterio rate coefficient
(KIE=$k_H/k_D$). Whereas for CH$_3$Cl an inverse KIE was determined
(i.e. KIE$<$1), the KIE was found to become increasingly more normal (KIE$>$1)
as the extend of methyl-substitution in the neutral reactant is increased. For
the reaction of BrO$^-$ with the fully methylated neutral species, an about a
factor of three larger reaction rate coefficient was measured for
(CH$_3$)$_3$CCl as compared to its deuterated form (CD$_3$)$_3$CCl. Villano et
al. referred to a marked effect of deuteration on the vibrational dynamics
near the respective transition states of the S$_{\rm N}$2 and the E2 mechanism
\cite{hu1996:jacs}. They argued that the E2 pathway is a minor channel for the
small neutral reactant CH$_3$Cl, whereas it becomes gradually more important
with increasing methylation of the neutral, which sterically hinders and
finally impedes nucleophilic substitution.

% example 3: POCl3-
The oxidation of the trichlorooxyphosphorus anion POCl$_3^-$, which occurs in
combustion flames, has been recently studied by Kerkines {\it et al.}
\cite{kerkines2010:jcp}. Despite its very low rate coefficient of only around
1x10$^{-14}$ cm$^3$/s (at 300K), such oxidations can change the chemistry of
flames due to the high abundance of O$_2$. The authors employed a turbulent
ion flow tube, where orders of magnitude higher neutral gas densities can be
applied as compared to the conventional laminar flow tubes
\cite{arnold2000:jpca}. They measured the rate coefficient over the range of
300-626\,K and found it to increase slowly with increasing temperature. An
Arrhenius fit yielded an activation energy of 50\,meV. Since the oxidation is
exothermic by about 1.8\,eV, the presence of an intermediate potential barrier
was concluded. Examination of the reaction pathways at different levels of
molecular orbital theory led to the proposal of a multistep reaction
mechanism. It involves product formation via the transformation of the
entrance channel ion-dipole complex into a four-membered P $\cdots$ O-O
$\cdots$ Cl ring transition state, the highest point on the potential energy
surface.

% example 4: CO3- + SO2
The highest temperature studies of ion-molecule reactions under fully
thermalised conditions have been undertaken in a high temperature flowing
afterglow apparatus at up to 1800K \cite{viggiano2001:agc}. For an
anion-molecule reaction, 1440K has been reached for CO$_3^-$ + SO$_2$
$\rightarrow$ SO$_3^-$ + CO$_2$ \cite{miller2006:jcp}. The main driving force
for these technically challenging experiments is to model the chemical
environment of the earth's ionosphere and other planetary atmospheres at high
altitude. However, these studies are also interesting from the reaction
dynamics point of view. At these high temperatures even small molecules carry
significant amounts of rotational and vibrational excitation. It can be
explored how energy supplied in different forms - translational or internal -
affects the reaction rate. For the reaction of CO$_3^-$ with SO$_2$, Miller et
al. found by comparison with drift tube data that the total energy alone
controls the reactivity \cite{miller2006:jcp}. They concluded that the
independence of the rate coefficient on the form of energy implies that the
reaction is governed by long lived intermediates in which energy equilibrates,
even at the very high temperatures.

%%%%%%%%%%%%%%%%%%%%%%%%%%%%%%%%%%%%%%%%%%%%%%%%%%%%%%%%%%%%%%%%%%%%%%%%%%%%%%

\subsection{Integral cross sections from guided ion beams}

% GIB technique: guiding the ion
Guided ion beam (GIB) studies are the method of choice for precise
measurements of integral reaction cross sections of ion-molecule reactions for
collision energies of the millielectronvolt to the tens of electronvolt range.
These measurements also allow for the determination of the opening and
competition of different reactive channels in ion-molecule collisions
\cite{armentrout2000:ijm}. Teloy and Gerlich introduced GIBs into gas-phase
chemistry in pioneering experiments in Freiburg in the 1970s
\cite{teloy1974:cp}. In this technique mass selected ions are passed into a
long radio frequency (rf) multipole ion guide - typically an octupole - with a
selected kinetic energy, controlled by the dc potential difference between the
ion source and the ion guide. Importantly, the guiding multipole electric rf
field contains the ions radially, while at the same time minimising alteration
of their kinetic energy. The latter is crucial to the technique and impedes
the use of quadrupole rf fields: Micro motion of the ions driven by the
oscillating field results in energetic collisions with buffer and reaction
gas. These effects are referred to as radio frequency heating and are
minimised in a multipole rf field \cite{gerlich1992:adv}. The created
effective potential guides the ions through a scattering cell located at the
centre of the long guide.

% GIB technique: adding the neutral collision partner
The neutral reaction partner is introduced into the scattering cell - which
may be temperature-variable \cite{levandier1997:rsi} - at a well characterised
density. Ions collide with the neutrals as they pass through the scattering
cell and undergo chemical reactions. Note that the multipole guide also
contains the ionic reaction products. All ions are mass analysed when they
reach the end of the guide, typically with a quadrupole mass
selector. Counting the number of reactant and product ions allows one to
determine absolute integral cross sections for ion-molecule reactions and
collision induced dissociation. These measurements can be done as a function
of the relative kinetic energy of the reactants over an extended range of
energies from about 0.1 to hundreds of eV. GIB measurements achieve high
resolution since they ensure single collision conditions in contrast to drift
tubes and employ better defined, narrow ion velocity distributions
\cite{deturi1997:jpc}. At the same time the range of relative collision
energies is extended since ions are radially contained and can not get lost by
drifting to the walls of the tube. To further minimise the energy spread of
the reactants, sophisticated ion sources such as flow tubes and multipole
traps are used \cite{haufler1997:jpc,deturi1997:jpc,gerlich2004:jams}.

% GIB for cations
The bulk of GIB studies have focused on positively charged ions. Negative ion
chemistry has been investigated in particular by the group of Kent Ervin, with
special emphasis on nucleophilic substitution. To uncover reaction dynamics
and mechanisms, the group heavily employs ab initio calculations and density
functional theory in the interpretation of their measurements. The high
resolution of GIB measurements allows to reveal the threshold behaviour of
endoergic reactions \cite{rempala2000:jcp} and of exothermic reactions, which
feature an intermediate potential barrier \cite{deturi1997:jpc}. The wide
tunability of the relative collision energy enables studies of the opening and
interplay of reaction mechanisms leading to different ionic products. Absolute
integral cross sections of competing nucleophilic substitution and abstraction
reactions were obtained \cite{angel2001:jpc,angel2002:jacs}. Collision
induced dissociation of anion-dipole complexes accesses the dynamics at the
transition state \cite{akin2006:jpc}. In an advanced mode of operation, GIB
experiments may be run in a pulsed manner and the dwell time of the ionic
reaction products in the long guide may be used to analyse their axial kinetic
energy distribution. This allows to some degree to distinguish between forward
and backward scattering for a more direct insight into ion-molecule reaction
dynamics \cite{haufler1997:jpc,angel2003:jacs}.

% GIB for H-
Haufler, Schlemmer and Gerlich carried out a pioneering GIB study for anions,
investigating the fundamental molecular reaction H$^-$ + D$_2$ $\rightarrow$
D$^-$ + HD and its isotopic variant D$^-$ + H$_2$ $\rightarrow$ H$^-$ + HD
\cite{haufler1997:jpc}. Integral cross sections have been obtained as a
function of collision energy between 0.1 and 10\,eV translational energy in
the centre-of-mass frame. They show an onset at around 0.3\,eV, a maximum at
around 1\,eV and a decrease at larger collision energies. Barriers of 350 and
330\,meV were deduced for H$^-$ + D$_2$ and D$^-$ + H$_2$, respectively. The
decrease of the cross section was attributed to the competition with
collisional electron detachment of the reactant anion. Interestingly, it was
found that the integral cross section of the heavy ion colliding with the
light molecule is larger by a factor of two, reaching 2.7\,\AA$^2$ at its
maximum. Axial product time-of-flight distributions have been measured by
recording arrival times and converted to differential cross sections
d$\sigma$/d$v_p$ ($v_p$ denotes the velocity component of the products along
the guide). Preferred forward scattering was found for low collision energies
just above threshold. By variation of the depth of the effective potential via
the rf amplitude, which probes the transversal velocity distribution, the
authors concluded that internal excitation of the product molecule is not
significant at these energies. Thus they were able to deconvolute
d$\sigma$/d$v_p$ to obtain angle-differential cross sections
d$\sigma$/d$\theta$. These could be cross-calibrated against crossed-beam data
\cite{mueller1996:jpb} (see section \ref{conv-crossed-beams:sect}). The
angle differential cross sections were found to be quite similar for the two
variants of the reaction, despite the big isotope effect observed in the
integral cross section. The authors concluded that the region of the potential
energy surface probed in the exit channel has to be similar and the increased
reactivity of D$^-$ + H$_2$ has to stem from dynamics in the entrance channel.

% GIB for SN2 reactions of anions
DeTuri {\it et al.} investigated the symmetric S$_{\rm N}$2 reaction Cl$^-$ +
CH$_3$Cl $\rightarrow$ ClCH$_3$ + Cl$^-$ \cite{deturi1997:jpc} by isotopic
labelling. For this system, a translational energy threshold of 2\,eV had been
determined 10 years before in a pioneering and well-recognised drift tube
measurement by the Bierbaum group \cite{barlow1988:jacs}. Based on that
finding, anionic attack on the chlorine side of chloromethane (``frontside
attack'') had been proposed as the reaction mechanism (potential barrier
2.0\,eV), and subsequently been backed by theoretical studies
\cite{deng1994:jacs,glukhovtsev1996:jacs2}. In contrast, DeTuri {\it et al.}
determined a translational energy threshold of 470\,$\pm$\,160\,meV in their
GIB experiment. The authors argued that the previous measurement had been
affected by the skewed reactant kinetic energy distribution in the drift tube
as well as by collisions with its high density helium buffer gas during the
lifetime of the entrance channel complex. The adjusted threshold, however,
energetically excludes frontside attack as the reaction mechanism. On the
other hand the conventional S$_{\rm N}$2 back-side attack mechanism, with
inversion of the carbon centre, features a potential barrier height of only
around 120\,meV, substantially lower than the measured threshold by DeTuri et
al.

% new theory on SN2
This puzzle prompted a lot of studies on the theoretical side. Quasiclassical
trajectory simulations by the Hase group both on an analytical potential
energy surface \cite{mann1998:jpc} and with the ab initio direct dynamics
method \cite{li1999:jacs} as well as reduced-dimensionality quantum dynamical
calculations \cite{hennig2005:pccp,hennig2004:jcp} were undertaken. It was
confirmed that frontside attack is not the reaction mechanism and backside
attack prevails. However, in contrast to the conventional S$_{\rm N}$2
mechanism, it was found that only a very small fraction of the reactive
trajectories are indirect, with transient trapping in the ion-dipole
wells. This is assigned to poor coupling of the provided translational
excitation to internal degrees of molecular freedom. A direct pathway of
product formation, on the other hand, seems to feature strict dynamical
constraints for low collision energy such as a very restricted geometry of
approach for passing over the minimum energy barrier.

\begin{figure}[tb]
\center
\includegraphics[width=\columnwidth]{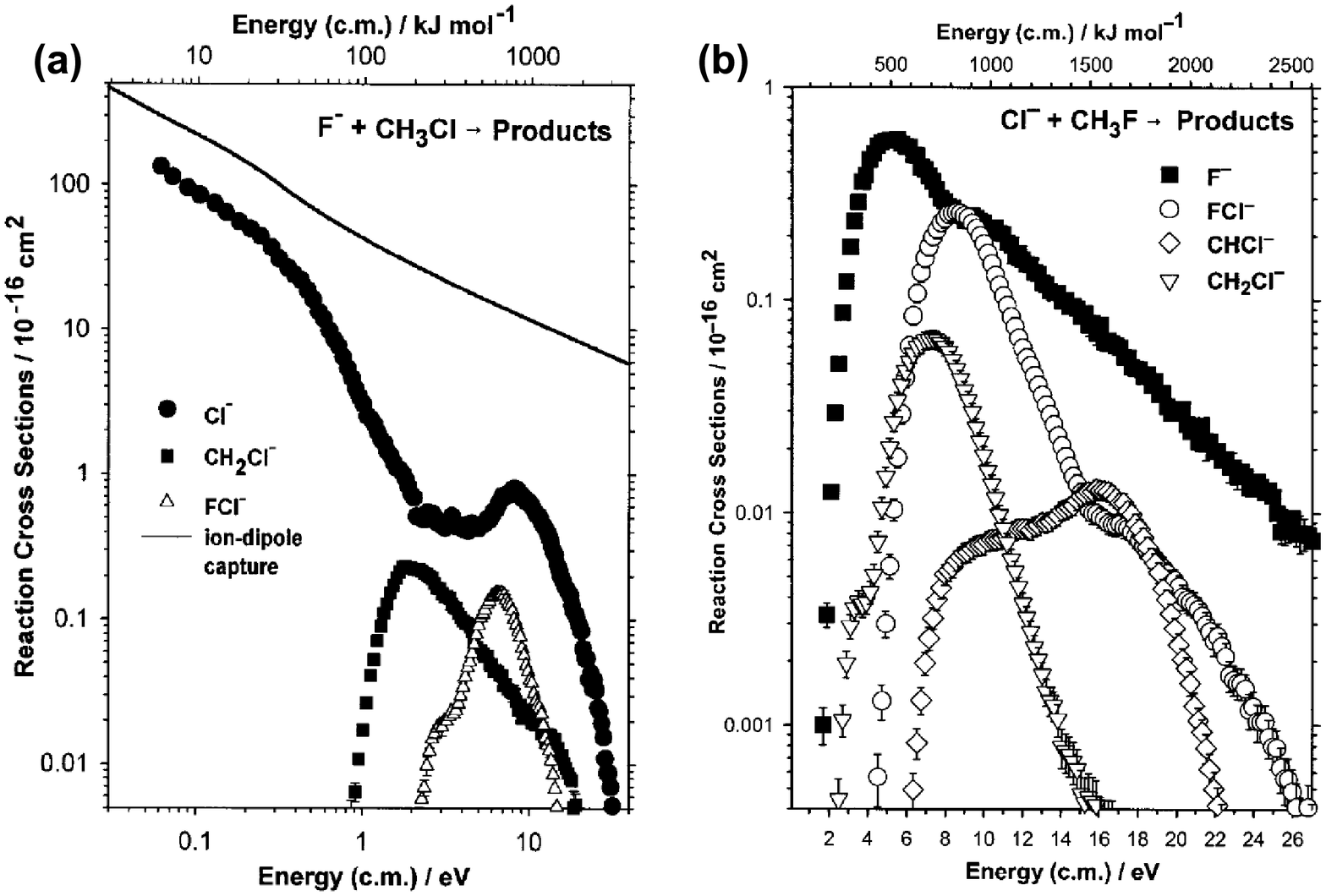}
\caption{Absolute integral cross section from a GIB study of F$^-$ +
  CH$_3$Cl (a) and the inverse reaction Cl$^-$ + CH$_3$F (b) for
  different product ions as a function of the relative collision
  energy. The calculated capture cross section is shown as solid
  line. Reprinted with permission from \cite{angel2001:jpc} and
  \cite{angel2002:jacs}. Copyright (2001), American Chemical Society.}
\label{angelervin}
\end{figure}

% SN2 GIB by Angel and Ervin
The power of the GIB technique is nicely demonstrated in complementary cross
section studies of the exothermic S$_{\rm N}$2 reaction F$^-$ + CH$_3$Cl and
its counterpart, the endothermic reverse reaction Cl$^-$ + CH$_3$F
\cite{angel2001:jpc,angel2002:jacs}. The data from Angel and Ervin are
reproduced in Fig.\ \ref{angelervin}a and b. As can been seen, the measured
reaction cross sections range over an impressive dynamic range from more than
10$^{2}$\,\AA$^2$ down to 5$\times$10$^{-4}$\,\AA$^2$. The exothermic S$_{\rm
  N}$2 reaction leading to Cl$^-$ product formation is most efficient at the
lowest centre of mass collision energies, while its cross section decreases
rapidly over the range 0.1-2\,eV. The reason for this behaviour derives from
the S$_{\rm N}$2 reaction mechanism. A more impulsive collision allows for
less alignment of the reaction partners along the backside attack coordinate
by their ion-dipole interaction, i.e. along the near collinear reaction
path. Fig.\ \ref{angelervin}b shows that the endothermic S$_{\rm N}$2 reaction
leading to F$^-$ product formation features an energy threshold as
expected. Angel and Ervin determined a threshold energy of 1.88\,eV, which is
0.54\,eV in excess of the reaction endothermicity.

%%% Figure

% interpretation of the results
Inspecting the pseudo-collinear PES, the authors interpreted their finding in
terms of the Polanyi rules \cite{polanyi1972:acr}. An early transition state
in the exothermic direction for the collinear reaction pathway renders
translational energy efficient in promoting the reaction. This accounts for
the high cross section in excess of 100\,\AA$^2$ for low collision
energies. After translational passage over the barrier trajectories need to
pass the ``bend'' along the reaction coordinate to form products. Hence much
of the reactant translation is expected to be converted into excitation of the
C-Cl stretch vibration of the product molecule. Microscopic reversibility
suggests in turn that vibrational excitation of CH$_3$Cl is needed in the
endothermic direction to efficiently pass the initial tight bend on the PES
before surmounting the late transition state. Translational activation such as
in the GIB experiment should hence be inefficient at promoting the endothermic
reaction. This is in accordance with the finding of an excess barrier by Angel
and Ervin and a low cross section of maximal 0.6\,\AA$^2$. Fig.\
\ref{angelervin} shows that at higher collision energies other reaction
channels open up and compete with nucleophilic substitution. For the
exothermic direction (Fig.\ \ref{angelervin}a), both proton transfer and
chlorine abstraction reaction were observed. At collision energies above
2\,eV, the Cl$^-$ cross section was found to increase again. Dissociation of
the CH$_2$Cl$^-$ and FCl$^-$ reaction products was proposed as an
explanation. For the endothermic direction (Fig.\ \ref{angelervin}b),
methylene and fluorine abstraction as well as a subsequent threebody
dissociation were observed. Some of these reactions exhibit dual rising
features. Reaction mechanisms have been proposed based on molecular structure
calculations.

% summary on GIB and dynamics
While GIB techniques are strong at determining integral reaction cross
sections, the very detailed and insightful study by Angel and Ervin also
clearly shows their limitations. An experimental method is needed, which can
provide more direct insight into the reaction dynamics and is capable of
dissecting the partitioning of energy into translational and internal degrees
of molecular freedom (see Section \ref{crossed-beams:sect}).

% differential measurements using GIB
As first demonstrated by Gerlich \cite{gerlich1992:adv} and already mentioned
above, GIB methods can in fact provide to some degree more direct insight into
the reaction dynamics by recording the arrival time of the product ions. Angel
and Ervin applied this technique in a study of the exothermic S$_{\rm N}$2
reaction Cl$^-$ + CH$_3$Br \cite{angel2003:jacs}. They reported the relative
product velocity distribution along the axis of the ion guide for a series of
relative collision energies between 0.1 and 4.0\,eV. At 0.1\,eV, the S$_{\rm
  N}$2 product velocity was found to be symmetrically distributed around the
centre of mass velocity and to agree well with the prediction of a statistical
phase space theory (PST) model. This can be regarded as strong sign that the
reaction is complex-mediated. At 0.25\,eV, the velocity distribution becomes
asymmetric with a preference for forward scattering of the CH$_3$Cl
product. This shows that the reaction becomes more direct, with reactant
interaction on a time scale less than the rotational period of the complex. At
0.5\,eV, backward scattering arises and becomes dominant above 1.0\,eV. The
authors interpreted this as introduction of a direct rebound mechanism.

%%%%%%%%%%%%%%%%%%%%%%%%%%%%%%%%%%%%%%%%%%%%%%%%%%%%%%%%%%%%%%%%%%%%%%%%%%%%%%

\subsection{Low-temperature reactions in jets and traps}

% interest in low temperature
At lower and lower temperatures the dynamics of ion-molecule reactions become
more sensitive to details of the interaction potential and to quantum
dynamical effects such as tunnelling. Changes in the rate coefficients at low
temperature also affect the abundance of ions in atmospheric and interstellar
plasmas, which are often characterised by temperatures well-below room
temperature. This has raised substantial interest in low-temperature studies
of ion-molecule reactions.

% supersonic studies
Very low temperatures, down to about 0.1\,K, are reached in free jet
supersonic flows employing adiabatic expansion \cite{smith1998:irpc}.
However, the complex cooling process that is induced by the transient flow
dynamics of an expanding free jet does lead to temperature disequilibria for
the different molecular degrees of freedom and non-thermal rotational state
distributions \cite{zacharias1984:jcp,belikov1998:cp}. Despite these
difficulties supersonic jets have been used to study a few reactions of
cations, produced by electron impact or laser ionisation. Two examples are the
study of vibrational quenching of NO$^+$($\nu$=1) via helium by complex
formation below 3\,Kelvin \cite{hawley1991:jcp} and the reopening of the
bimolecular C$_2$H$_3^+$ channel in the hydrogen transfer of H$_2$ to
C$_2$H$_2^+$ for low temperatures, which is attributed to tunnelling of a
collision complex through the reaction barrier \cite{hawley1992:jcp}. An
extension of this technique to negative ion reactions is not straight forward
and has not been carried out up to now.

% CRESU studies
Uniform expansions from precisely designed Laval nozzles can overcome some of
the difficulties of free jet flows by maintaining parallel stream lines at
constant Mach number \cite{smith1998:irpc}.  Constant high densities are
reached at a fixed temperature in the travelling frame over the entire flow,
which makes equilibration of the molecules degrees of freedom more likely. By
terminating the supersonic expansion through parallelisation of the stream
lines in the nozzle, the exceedingly low temperatures of free expansions are
unfortunately lost. This CRESU (a French acronym for Cin\'{e}tique de
R\'{e}action en Ecoulement Supersonique) technique is applicable to negatively
charged ions, created by dissociative electron attachment in the isentropic
flow.

% SN2 with CRESU
CRESU was employed by Le Garrec {\it et al.} to study the S$_{\rm N}$2 reaction
Cl$^-$ + CH$_3$Br from 180 down to 23\,K \cite{legarrec1997:jcp}. Combining
their results with previous measurements obtained from other experimental
techniques, the authors demonstrated a dramatic increase in the rate
coefficient by a factor of 400 upon reduction of the temperature from 500 to
23\,K. A quantum scattering method, the ``rotating bond approximation'' has
been employed in this work. It was found that at the lowest temperatures the
reaction follows the potential energy profile with no activation energy
present. The excitation of intermolecular bending modes of the transition
state then induces an activation barrier for increasing temperature.

% reactions in traps
With the exploration of radio-frequency heating in guided ion beams, it became
clear that for the efficient cooling and complete thermalisation of molecular
ions with a buffer gas, trapping in a quasi field-free environment is
indispensable \cite{gerlich1992:adv}. Gerlich pioneered the development of
cryogenic storage devices based on electric multipole fields, which are
capable of cooling molecular ions to a few Kelvin in all degrees of molecular
freedom. The most popular design is the 22pole ion trap \cite{gerlich1995:ps},
which is nowadays used in a wide range of applications \cite{wester2009:jpb}.

% coldest anion-molecule reaction
In such a temperature-variable 22-pole ion trap, the lowest-temperature for an
anion-molecule reaction up to now has been achieved in a recent experiment in
our group \cite{otto2008:prl}. The specific system under investigation was the
proton transfer reaction to the negatively charged amide ion
\begin{equation}
{\rm NH}_2^- + {\rm H}_2 \rightarrow {\rm NH}_3 + {\rm H}^-.
\label{nh2:reaction}
\end{equation}
For this reaction a room temperature reaction probability, given by the ratio
of the measured rate coefficient to the calculated Langevin rate coefficient,
of about 2\% had been found many years ago \cite{bohme1973:jcp}. Thus, in 98\%
of the collision events the reactants are reproduced and no products are being
formed. This occurs despite the fact that the reaction is exothermic and also
the intermediate potential barrier lies below the energy of the entrance
channel. It shows that for this reaction the reaction dynamics at short range
are very important.

%%% Figure
\begin{figure}[tb]
\center
\includegraphics[width=\columnwidth]{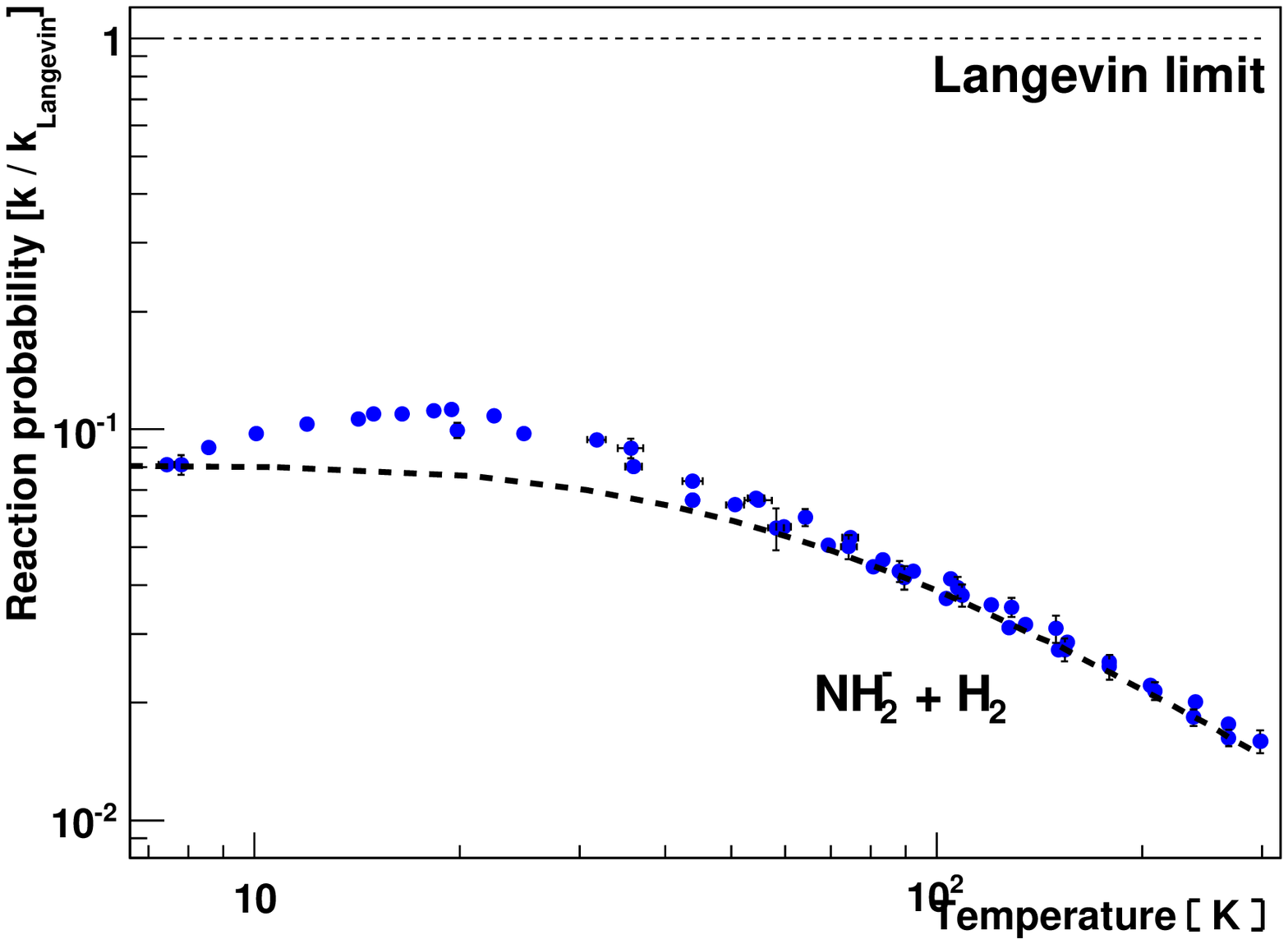}
\caption{Reaction probability, given by the ratio of the measured rate
  coefficient and the constant Langevin rate coefficient, for the reaction of
  NH$_2^-$ with H$_2$ (Eq.\ \ref{nh2:reaction}) as a function of temperature
  \cite{otto2008:prl}. The data show that the probability for reaction
  increases with decreasing temperature but stays well below the Langevin
  limit. Below 20\,K the data show an unexpected decrease of the reaction
  probability that can not be explained by a classical statistical model
  (dashed line).}
\label{reactions:fig} 
\end{figure}

% nh2- + H2
At lower temperatures the probability for reaction (\ref{nh2:reaction})
increases strongly \cite{otto2008:prl}. The data are shown in the right panel
of Fig.\ \ref{reactions:fig}. At 20\,K the probability has increased by a
factor of six. This increase is a manifestation of the complex-mediated
reaction dynamics \cite{troe1994:far}: the intermediate NH$_4^-$ complex,
which is transiently formed during a collision that surmounts the centrifugal
barrier, has a longer lifetime with respect to decay back to reactants at
lower temperatures, because the number of available decay channels
decreases. The probability to cross the intermediate potential barrier and
form products, however, remains approximately constant. Therefore the overall
probability to react increases. The observed decrease of the reaction
probability for temperatures lower than 20\,K (see Fig.\ \ref{reactions:fig})
can not be explained within the classical dynamics picture of a
complex-mediated reaction mechanism. Instead it is expected to represent a
signature of quantum mechanical reaction dynamics in low temperature
ion-molecule reactions.

%%%%%%%%%%%%%%%%%%%%%%%%%%%%%%%%%%%%%%%%%%%%%%%%%%%%%%%%%%%%%%%%%%%%%%%%%%%%%%
%%%%%%%%%%%%%%%%%%%%%%%%%%%%%%%%%%%%%%%%%%%%%%%%%%%%%%%%%%%%%%%%%%%%%%%%%%%%%%
%%%%%%%%%%%%%%%%%%%%%%%%%%%%%%%%%%%%%%%%%%%%%%%%%%%%%%%%%%%%%%%%%%%%%%%%%%%%%%

\section{Probing the collision complex}

In the following section we discuss experiments that investigate the transient
complex that the reactants form along the path to reaction
products. Experimentally, this is either achieved by analysing ternary,
complex-forming reactions, by studying unimolecular decay of anion-molecule
compounds or by applying direct time-resolved spectroscopy, as detailed below.

%%%%%%%%%%%%%%%%%%%%%%%%%%%%%%%%%%%%%%%%%%%%%%%%%%%%%%%%%%%%%%%%%%%%%%%%%%%%%%

\subsection{Three-body association and dissociation of reaction intermediates}

Average lifetimes of transient entrance channel complexes formed upon
collision of the reaction partners can be determined indirectly from ternary
rate coefficients for their formation and collisional stabilisation. For
reactions of cations with neutral molecules, the extraction of complex
lifetimes from ternary association has a long history using flow and drift
tubes at high pressure. For reactions of anions, work has focused on the
symmetric S$_{\rm N}$2 reaction Cl$^-$ + CH$_3$Cl, specifically the formation
and unimolecular decay time of the metastable [Cl$^-\cdots\,$CH$_3$Cl]$^*$
entrance channel complex \cite{li1996:jacs,mikosch2008:jpc}.

Ternary association is studied in a thermal environment, where the density of
the neutral reaction partner and a buffer gas is well defined and
controllable. It can be understood as a two-step process.
\begin{equation}
  \label{3body_equations}
  \begin{array}{c c c c c}
    & k_{f} n_{\rm CH_3Cl} & & k_{s} n' & \\
    \mathrm{Cl^- + CH_3Cl} & \stackrel{\rightharpoonup}{\leftharpoondown} &
    \mathrm{[Cl^-\cdots CH_3Cl]^*} & \rightarrow & \mathrm{[Cl^-\cdots CH_3Cl]}\\
    & k_{\Gamma} & & &\\
  \end{array}
\label{ThreeBody}
\end{equation}
In the first step collisions of the reactants Cl$^-$ and CH$_3$Cl form the
metastable ion-dipole complex under investigation, where the relative kinetic
energy is transiently transferred into internal energy. The complex is highly
excited with respect to its ground state and decays back to reactants with the
unimolecular rate k$_{\Gamma}$. Stabilisation of the complex can occur in a
second step if a third particle impact removes more internal excitation from
the complex than the initial relative translational and internal energy of its
constituents. This is only a small fraction of its internal excitation given
the binding energy of ground state [Cl$^-\cdots\,$CH$_3$Cl] of about 450\,meV
with respect to the asymptote. In subsequent collisions the internal energy is
thermalised with the environment. Re-excitation of the complex above the
asymptote for dissociation is then strongly suppressed by a Boltzmann
factor. Conditions are arranged such that the stabilisation step occurs with a
high probability by the buffer gas, which is provided in large excess as
compared to the reaction partners. At the same time the low-pressure limit of
the steady state approximation has to be valid, which practically means that
it has to be ensured that stabilisation of the complex is a rare event. Then
the rates for the association and stabilisation step in (\ref{ThreeBody}) are
well defined and the overall rate of [Cl$^-\cdots\,$CH$_3$Cl] formation is R =
k$_3$ n$_{\rm CH_3Cl}$n, where k$_3$ is the ternary rate coefficient k$_3$ =
k$_f$k$_s$/k$_{\Gamma}$. The ternary rate coefficient is experimentally
determined by measuring R as a function of the densities of the reactant
(n$_{\rm CH_3Cl}$) and the buffer gas (n). The rate coefficients k$_f$ and
k$_s$ describe barrier-free ion-molecule reactions; the method relies on the
assumption that these rate coefficients are very close to the capture-limited
values. With this, the unimolecular dissociation rate can be extracted, the
inverse of which is the average lifetime of the excited collision complex
[Cl$^-\cdots\,$CH$_3$Cl]$^*$.

Li {\it et al.} studied ternary association with high pressure mass
spectrometry and obtained an average lifetime of 12 to 16\,ps for
[Cl$^-\cdots\,$CH$_3$Cl]$^*$ at a fixed temperature of 296\,K
\cite{li1996:jacs} (see also \cite{mikosch2008:jpc}). In a recent experiment
in our group, a 22-pole radiofrequency ion trap has been employed, which
provides the advantage of temperature-variability and long interaction times
\cite{mikosch2008:jpc}. This allows to study the complex lifetime as a
function of reactant collision temperature down to 150\,K, at which point the
vapour pressure of the neutral reactant becomes too low to be bearable. We
have found a strong inverse temperature dependence of the average lifetime of
the transient [Cl$^-\cdots\,$CH$_3$Cl]$^*$ complex in disagreement with a
simple statistical model. Longer lifetimes of the entrance channel complex at
lower temperatures give S$_{\rm N}$2 reaction systems more time to exploit the
available phase space, randomise energy, cross the central reaction barrier
and finally form reaction products. This substantiates further that enhanced
lifetimes of the transient entrance channel complex are likely to underlie the
strong inverse temperature dependence of rate coefficients for exothermic
S$_{\rm N}$2 reactions with a submerged barrier.

Unimolecular dissociation of the metastable reactant complex into products
\begin{equation}
[{\rm X}^-\cdots\,{\rm CH}_3{\rm\,Y}]^* \rightarrow {\rm Y}^- + {\rm CH}_3{\rm\,X}
\label{UnimolecDiss}
\end{equation}
was studied in a series of experiments on asymmetric S$_{\rm N}$2 reactions in
the Bowers group \cite{graul1991:jacs,graul1994:jacs,graul1998:jacs}. Since
entrance channel complexes were prepared, the formation of products involves
at least one crossing of the intermediate reaction barrier. Relative kinetic
energy distributions for the released products were recorded by means of ion
kinetic energy spectroscopy. The measurements were compared to statistical
phase space theory. Somewhat controversial in these decomposition experiments
is the initial state of the metastable reactant complexes under
investigation. They have to have sufficient lifetime to survive the transport
from the high pressure ion source to the field-free high vacuum region of the
employed mass spectrometer and still undergo unimolecular dissociation. This
renders partial collisional stabilisation in the ion source likely (see also
\cite{seeley1997:jacs}). The kinetic energy release distributions for the
investigated S$_{\rm N}$2 reactions all peak at zero relative kinetic energy
of the products and rapidly decrease for increasing energy. Interestingly,
despite the long complex lifetimes the experimental distributions are much
narrower than calculated ones based on phase space theory. This means that
less energy is partitioned to relative translation of the dissociating
products than predicted for a statistical redistribution of energy. The phase
space calculation was brought into agreement with the experimental results
only if a significant amount of energy was made unavailable for energy
redistribution. This effect remains if partial collisional stabilisation is
taken into account. The energy missing in translation has to be trapped in
internal degrees of freedom of the reaction products. The authors argued that
significant rotational excitation is not present and that hence the neutral
reaction product has to be vibrationally hot. Surprisingly, the same result
was found for ``simple'' S$_{\rm N}$2 reactions of methylhalides
\cite{graul1994:jacs} as well as for S$_{\rm N}$2 reactions involving reaction
partners with many more internal degrees of freedom
\cite{graul1998:jacs}. While for the methylhalide reactions this observation
is in agreement with classical trajectory studies \cite{wang1994:jacs}, it
challenges the notion that statistical theories become valid for more complex
S$_{\rm N}$2 reactions with extended complex lifetimes, whereas
non-statistical dynamics is restricted to small or highly energised systems
with short complex lifetimes on the order of tens of picoseconds
\cite{craig1999:jacs,laehrdahl2002:ijms}. For extended complex lifetimes, it
was speculated that tunnelling through the central barrier might be an
alternative way of product formation, leading to a reduced effective barrier
height \cite{seeley1997:jacs}.

Dissociation of reaction complexes is ideally studied upon state-specific
excitation of an initially cold complex. For Cl$^-\cdots$\,CH$_3$Br it was
found that excitation of high-frequency intramolecular vibrations in the
CH$_3$Br moiety by a CO$_2$ laser leads exclusively to the product formation
Br$^-$ + CH$_3$Cl \cite{tonner2000:jacs}. This was predicted for this system
by classical trajectory calculations from Hase and coworkers
\cite{wang1994:jacs}, which demonstrated an enhancement of central barrier
crossing upon selective excitation. Similarly, excitation of the doubly
degenerate C-H stretch modes in the same complex induces central barrier
crossing and the formation of Br$^-$ products \cite{ayotte1999:jacs}. Craig et
al. created the S$_{\rm N}$2 intermediate
[CF$_3$CO$_2$CH$_3\cdots$\,Cl$^-$]$^*$ in a highly vibrationally excited state
by means of a precursor exothermic association reaction
\cite{craig1998:jacs}. This resulted in an at least four-fold enhancement of
the branching ratio for the S$_{\rm N}$2 reaction pathway as compared to
unexcited complexes. The observation is in disagreement with statistical RRKM
calculations and was interpreted as manifestation of a bottleneck for energy
transfer between intra- and intermolecular modes on the timescale of the
complex lifetime \cite{hase1994:sci}.

%%%%%%%%%%%%%%%%%%%%%%%%%%%%%%%%%%%%%%%%%%%%%%%%%%%%%%%%%%%%%%%%%%%%%%%%%%%%%%

\subsection{Time-resolved photoelectron spectroscopy}

Chemical reactions involve the nuclear motion from reactants to products as
well as the coupled structural and energetic transformation of molecular
orbitals. Ultrafast laser pulses allow to follow half-reactions in real time
by photoinitiating the dissociation of a transition state and probing the
evolution to products with a second photon in a pump-probe experiment
\cite{scherer1990:jcp,williamson1997:nat}. In time-resolved photoelectron
spectroscopy (TRPES) the probe laser generates free electrons by
photoionisation or photodetachment, whose kinetic energy and eventually
angular distribution is measured. Since TRPES is sensitive to both the
electronic configurations and the vibrational dynamics, it has been a
particularly successful tool for real-time insight into molecular
photodynamics (see recent reviews
\cite{stolow2004:chemrev,stolow2008:adv,mabbs2009:csr}).

In pioneering TRPES studies of negative ions, Neumark and coworkers
time-resolved the photodissociation of the I$_2^-$ anion
\cite{greenblatt1996:cpl,zanni1999:jcp}. An ultrashort laser pulse at 780\,nm
pumps ground state I$_2^-$ ($X ^2\Sigma_u^+$) to the first excited electronic
state ($A'$ $^2\Pi_{g,1/2}$). Following this excitation, I$_2^-$ dissociates
into ground state products I$^-$ + I($^2P_{3/2}$) with a 0.6\,eV kinetic
energy release. At variable time delay with respect to photoinitiation an
ultrashort probe pulse is employed, which produces a photoelectron spectrum
(PES). For a 260\,nm probe wavelength, two photoelectron bands are observed,
which asymptotically correspond to photodetachment of the I$^-$ photo-reaction
product. For short pump-probe delays, the photoelectron bands shift to smaller
kinetic energy for increasing delay, while at the same time their width
narrows (see also \cite{sanov2008:irpc}). This effect was attributed to the
change in character of the orbital from which the photoelectron is derived. By
tracing the transition from the molecular to atomic orbital, the PES uncovers
that the dissociation is complete within the first 320\,fs after
photoinitiation. Following dissociation, a subtle shift of the photoelectron
kinetic energy in the opposite direction is observed, which continues for
another 400\,fs. This behaviour was assigned to the interaction of the
separating fragments, in particular the polarisation-induced charge-dipole
attraction between the anion and the neutral atom. It corresponds to a shallow
well on the long-range part of the potential surface, which was characterised
in the TRPES experiment and determined to be 17\,meV deep
\cite{zanni1999:jcp}. Measured photoelectron angular distributions reveale
that the localisation of the excess charge on one of the atoms is only
complete after about 800\,fs \cite{davis2003:jcp}. The observed dynamics in
the exit channel of I$_2^-$ photodissociation becomes upon time-reversal
entrance channel dynamics for an I$^-$ + I collision.

Mabbs {\it et al.} \cite{mabbs2005a:jcp,sanov2008:irpc} extended this work to
the related system IBr$^-$. Here excitation at 780\,nm - to the lowest
optically bright excited electronic state - correlates to the second lowest
product channel, I$^-$ + Br($^2P_{3/2}$). Also here, a well is found in the
long range part of the potential, but compared to I$_2^-$ it is with about
60\,meV considerably deeper. Based on the result of a classical trajectory
calculation, the authors transformed the time-axis to the intermolecular
distance R - thus providing an ``image'' of the potential as a function of
reaction coordinate. Sheps {\it et al.} \cite{sheps2010:sci} showed very
recently that for the same photoexcitation the presence of a single solvent
molecule introduces a new product channel, the formation of I + Br$^-$. This
was traced back to a non-adiabatic transition to one of the lower-energy
electronic states driven by the solvent molecule, whose vibrational
temperature was found to play a critical role in the process.

%%% Figure
\begin{figure}[tb]
\center
\includegraphics[width=\columnwidth]{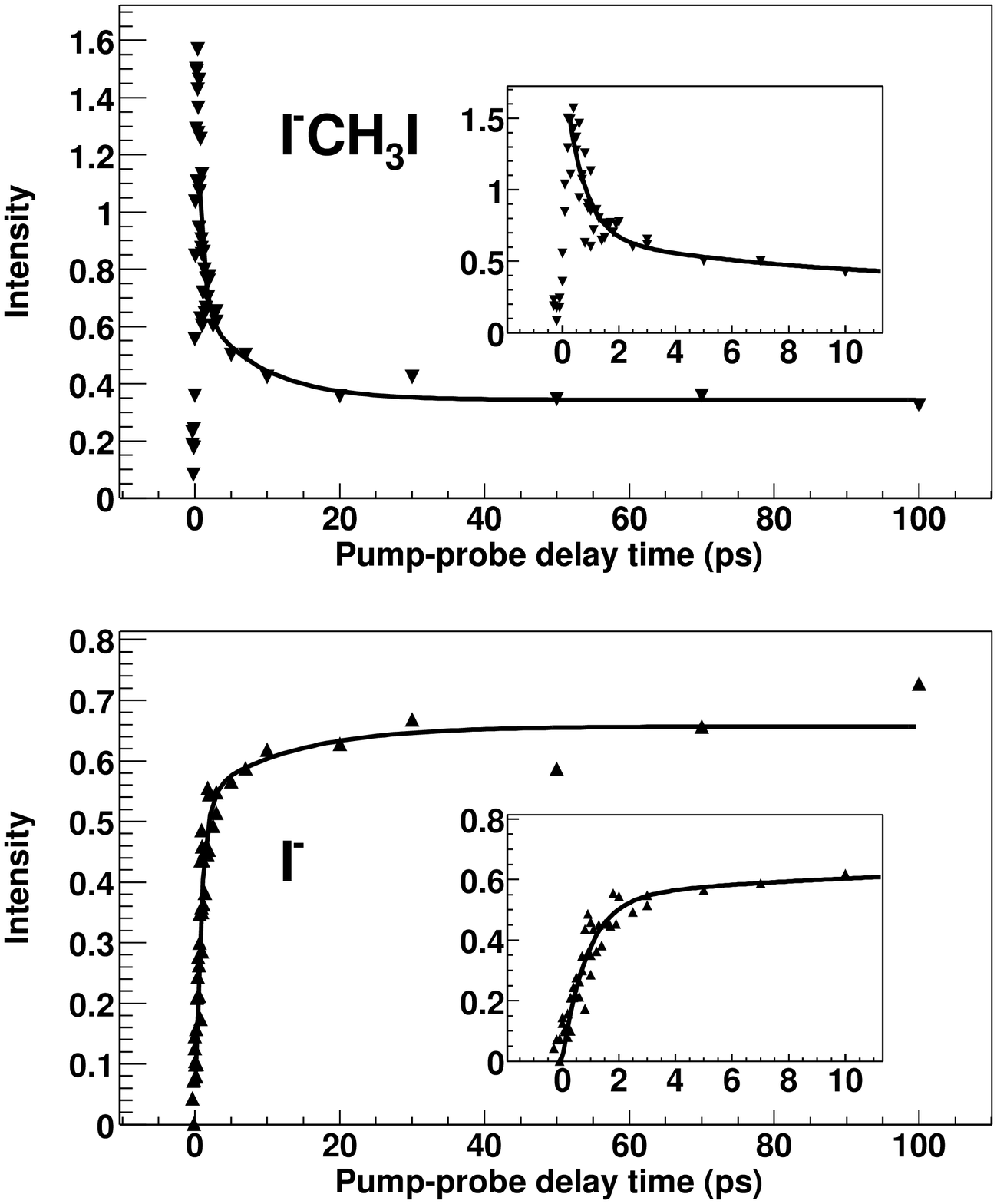}
\caption{Signal intensity of I$^-$CH$_3$I (upper panel) and I$^-$ (lower
  panel) in the femtosecond photoelectron spectra. Biexponential decay or
  growth curves, shown as solid lines, are found reproduce these signals
  \cite{wester2003:jcp}.}
\label{sn2-fpes:fig}
\end{figure}

One of the authors {\it et al.} performed the only time-resolved investigation
of a bimolecular S$_{\rm N}$2 reaction in the gas phase by studying I$^-$ +
CH$_3$I \cite{wester2003:jcp}. The reactants were derived from the precursor
ion-dipole-bound cluster I$_2^-\,\cdot\,$CH$_3$I. Start trigger for the
reaction is a femtosecond pump pulse, which dissociates the I$_2^-$
chromophore. Upon its photodissociation, the neutral iodine leaves the
cluster, while the S$_{\rm N}$2 reactants I$^-$ and CH$_3$I start to interact
due to their charge-dipole interaction. A probe pulse creates a PES, which
identifies reaction transients and products as a function of interaction
time. In the experiment two different pump photon energies have been employed,
which result in different kinetic energies of the I$^-$ reactant. At
$\lambda_{\rm pump}$ = 790\,nm, this leads to a relative kinetic energy
between I$^-$ and CH$_3$I of 0.15\,eV, where the interaction of CH$_3$I with
I$^-$ during dissociation has been neglected. The PES reveal that in this case
the dominant process is the production of the vibrationally excited entrance
channel complex I$^-\,\cdot\,$CH$_3$I on a time scale of 600\,fs. In contrast
for $\lambda_{\rm pump}$ = 395\,nm, the relative kinetic of the reactants
I$^-$ and CH$_3$I is 0.32\,eV (see Fig.\ \ref{sn2-fpes:fig}). Contributions of
I$^-\,\cdot\,$CH$_3$I and I$^-$ are now identified in the PES and can be
separated. The complex contribution increases rapidly after the pump pulse
with a time constant indistinguishable from the laser cross correlation of
200\,fs. Interestingly, it then shows an exponential decay on two different
time scales, which is mirrored in the I$^-$ contribution. Since the reactants
do not have sufficient energy to cross the central reaction barrier, this is
interpreted as decay of the complex back to reactants. The fast time constant
was determined to be 0.75\,ps, which compares to the vibrational period of the
I$^-\cdots\,$CH$_3$I stretching mode. It suggests fairly direct dissociation
dynamics in which the I$^-$ undergoes a quasi-elastic collision with CH$_3$I
before dissociation. The longer time constant of about 10\,ps indicates that
the complex is stabilised by energy flow from the reaction coordinate into the
modes of the complex. It is comparable to the lifetime of the
Cl$^-\,\cdot\,$CH$_3$Cl entrance channel complex determined in the collisional
stabilisation experiments featured above.

%%%%%%%%%%%%%%%%%%%%%%%%%%%%%%%%%%%%%%%%%%%%%%%%%%%%%%%%%%%%%%%%%%%%%%%%%%%%%%
%%%%%%%%%%%%%%%%%%%%%%%%%%%%%%%%%%%%%%%%%%%%%%%%%%%%%%%%%%%%%%%%%%%%%%%%%%%%%%
%%%%%%%%%%%%%%%%%%%%%%%%%%%%%%%%%%%%%%%%%%%%%%%%%%%%%%%%%%%%%%%%%%%%%%%%%%%%%%

\section{\label{crossed-beams:sect} Dynamics from differential scattering}

In the previous sections we have discussed measurements of the total cross
section or the thermally averaged rate coefficient and illustrated how
information on the reaction dynamics can be inferred from these
measurements. A much more direct approach to the intrinsic dynamics of
anion-molecule reactions is based on the measurement of the angle- and
energy-differential scattering cross section $d\sigma/d\Omega/dE$. For these
experiments collisions of atoms and molecules with well-defined momentum
vectors are prepared. Angle- and energy- or velocity-resolved detection
schemes are employed to obtain the differential cross section. This approach
of crossed-beam measurements of the differential scattering cross section is
widely used for neutral reactions \cite{casavecchia2000:rpp,liu2001:annu}. We
will discuss in the following two sections that it also proves to be valuable
for anion-neutral reactions.

%%%%%%%%%%%%%%%%%%%%%%%%%%%%%%%%%%%%%%%%%%%%%%%%%%%%%%%%%%%%%%%%%%%%%%%%%%%%%%

\subsection{\label{conv-crossed-beams:sect} Conventional crossed beam reactive scattering}

% conventional crossed beam scattering setup
The classical approach to crossed-beam reactive scattering is to cross two
reactant beams at 90$^\circ$ and use a rotatable detector to measure the flux
and the arrival time of products under a selected set of scattering angles in
the laboratory frame of reference. From these laboratory flux data the
differential cross section in the centre-of-mass frame of the reaction is
reconstructed, often using numerical simulations of the apparatus
function. This approach has been developed for neutral-neutral reactions and
has been used with great success to study the F + H$_2$ reaction (see e.g.\
\cite{lee1986:nl}). In combination with the Rydberg tagging technique to
measure the H atom product, the rotatable-detector setup is still the method
of choice for this reaction \cite{qiu2006:sci}. 

% rotatable detector for ion-molecule reactions
The same detector concept has been applied to study elastic, inelastic and
reactive channels in cation-molecule reactions (see Refs.\
\cite{futrell1992:adv,farrar1995:annu} for reviews of the early work on
cation-molecule reactive scattering). Here the main difference to
neutral-neutral scattering experiments is, naturally, the production of the
ion beam. Where the neutral beams are usually produced in supersonic
expansions with a well defined narrow range of velocities, the production of
ion beams with a narrow velocity spread represents a major experimental
challenge. For the study of chemical reactions the most interesting relative
collision energies in the centre-of-mass frame range from millielectronvolts
to a few electronvolts. Therefore ion beams have to be produced with a low
kinetic energy and a correspondingly small energy spread. This has been
achieved for the first time in the 1970s for continuous beams of cations
\cite{vestal1976:rsi}. A high degree of control over the electric fields in
the experimental setup, including contact potentials between different
materials, are important to achieve a low energy spread. Even then, the
Coulomb repulsion between the reactant ions in the beam ultimately limits the
energy resolution for a given reactant density and thus scattering rate. For
ion-molecule reactions this typically leads to a reduction in the ion reactant
density by many orders of magnitude in comparison with a neutral supersonic
beam. This is only to a small extend compensated by the higher detection
efficiency for product ions in a charged-particle detector when compared to
neutral products of neutral-neutral reactions which have to be ionised in the
detector. Note that once the control over the ion-beam energy is achieved in
an experimental setup, the relative collision energy of the crossed beams can
be continuously tuned by means of the ion acceleration potential. Such a
tunable scattering energy is much more difficult to achieve for
neutral-neutral reactions. There, different seed gases for the molecules in
the supersonic beams and different intersection angles between the supersonic
beams are required for the same purpose.

% early crossed-beam work on anions
It was only in the 1990s that precise low-energy crossed-beam measurements
with negative ions in a rotatable detector setup have been carried out
\cite{zimmer1995:jpb,farrar1995:annu}. In the years before the first detailed
scattering experiments on anion-molecule reactions have been carried out using
a complex multiparticle coincidence scattering setup. With this setup
reactions of heavy, mostly atomic, negative ions (F$^-$, Cl$^-$, Br$^-$,
I$^-$, S$^-$, CN$^-$) with light-weight hydrogen could be studied
\cite{barat1985:cp,fayeton1989:cp,brenot1994:cp,goudjil1994:cp}. For example
for the Cl$^-$ + H$_2$ collision, the kinematics of this combination have made
it possible to study collisions down to 6\,eV relative energy with an ion
kinetic energies of 110\,eV \cite{barat1985:cp}, which is significantly easier
to handle experimentally than ion beams with only few electronvolt kinetic
energy. Furthermore, under these conditions the kinetic energy of the neutral
product in the lab frame is large enough to be detected with a microchannel
plate detector. With this technique the different channels for the collision
of Cl$^-$ with H$_2$, reactive proton transfer, forming HCl + H$^-$, reactive
detachment, forming HCl + H + e$^-$, simple detachment, leading to Cl + H$_2$
+ e$^-$, and dissociative detachment, leading to Cl + H + H + e$^-$, could be
distinguished and their differential and absolute cross sections could be
measured as a function of the relative energy \cite{barat1985:cp}. Also some
information on the vibrational state population of the product molecule could
be inferred. From these data the similarity of the dynamics of reactive proton
transfer and reactive detachment could be deduced, where the branching between
the two channels depends on the coupling to an intermediate autodetaching
HCl$^-$ state. Different scattering dynamics was found for the simple
detachment channel, bearing similarity with anion-rare gas collisions.

% lindner work on H-
With the first crossed-beam measurements on anion-molecule reactions using
low-energy ion sources and rotatable detectors, relative collision energies
down to the millielectronvolt range and an energy resolution sufficient to
resolve product vibrational states became accessible. In a benchmark
experiment on the reaction of H$^-$ with D$_2$ Zimmer and Linder could
determine the vibrational state-to-state differential and integral scattering
cross section \cite{zimmer1992:cpl,zimmer1995:jpb}. The dominate forward
scattering of the D$^-$ product was attributed to a collinear reaction
mechanism that requires scattering at small impact parameter. Wider angular
distributions were observed for higher vibrational excitation of the HD
product. From the measured product kinetic energy spectra also rotational
state information could be extracted, studied in more detail for the inelastic
scattering in H$^-$ + H$_2$ collisions in the same laboratory
\cite{mueller1996:jpb}.

%%% Figure
\begin{figure}[tb]
\center
\includegraphics[width=\columnwidth]{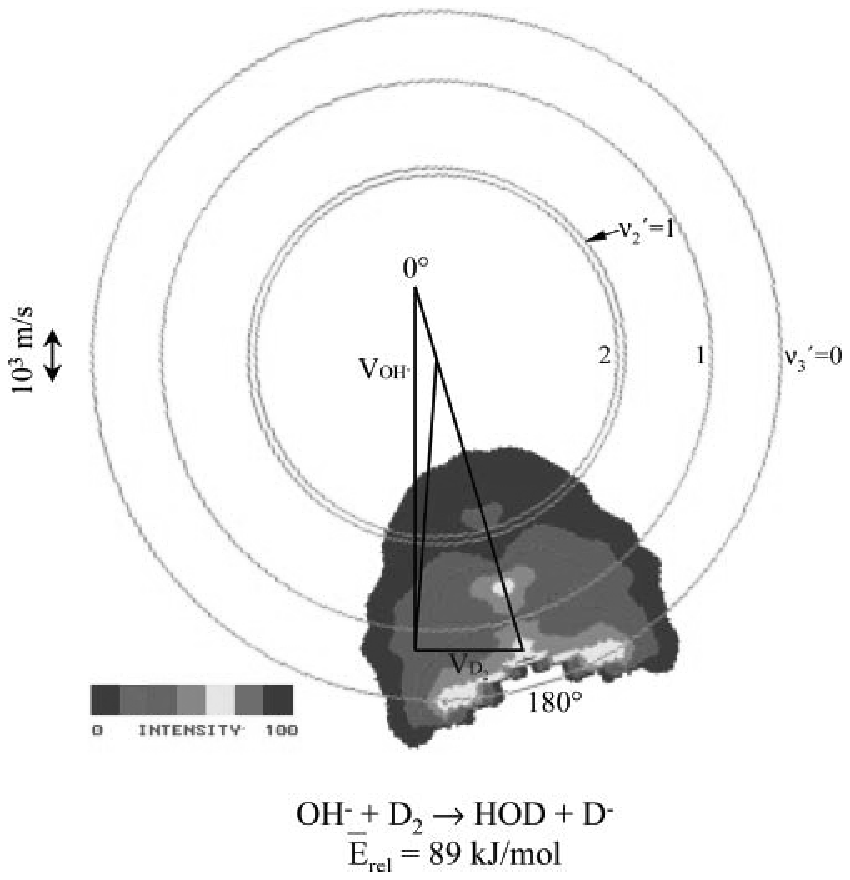}
\caption{Newton diagram and extracted product flux contour map for the
  reaction OH$^-$ + D$_2$ at 89\,kJ/mol relative energy. The orthogonal black
  lines indicate the velocities of the incoming beams. The point in the centre
  of the circles represents zero velocity in the centre of mass frame. Around
  this point the circles denote different product velocities. Smaller
  velocities indicate internal excitation of the HOD product molecule due to
  vibrational excitation of the bending mode (v$_3$) or the OD stretching mode
  (v$_2$). The measured flux shows that internal excitation, attributed to the
  bending mode of HOD, does occur during the reaction. Reprinted with
  permission from \cite{li2005:jpc}. Copyright (2005), American Chemical
  Society.}
\label{farrar-data:fig}
\end{figure}

% farrar O- and OH- work
Also in the 1990s, Farrar and coworkers started crossed-beam reactive
scattering experiments of atomic oxygen with small closed-shell molecules,
such as water, ammonia and hydrogen (see e.g.\
\cite{varley1992:jcp,levandier1992:jcp,farrar1995:annu}). For the O$^-$ +
H$_2$O reaction, which had already been studied before with crossed beams at
higher collision energy and with lower resolution \cite{karnett1981:cpl}, they
observed in the differential cross section both a direct and an indirect
reaction mechanism for the reactive channel of OH$^-$ formation
\cite{varley1992:jcp}. In addition they could study inelastic scattering via
the O$^-$(H$_2$O) reaction complex. More recently, non-reactive and reactive
collisions of OH$^-$ anions with D$_2$ molecules have been studied by the same
group \cite{lee2000:jcp,li2005:jpc}. In the isotopic exchange reaction
channel, the formation of OD$^-$ + HD, no energy dependence was found between
0.27 and 0.69\,eV relative collision energy \cite{lee2000:jcp}. In the
measured narrow angular distribution it was found that the reaction occurs
fast on the time scale of rotation of the OH$^-$(H$_2$) reaction complex. In a
study of the proton transfer reaction channel, forming HOD + D$^-$, also a
very narrow velocity distribution was extracted \cite{li2005:jpc}. The
reconstructed velocity distribution in the scattering plane is reproduced in
Fig.\ \ref{farrar-data:fig}. These data also indicate a direct and fast
reaction mechanism. The flux at smaller product ion velocities is caused by
vibrational excitation of the bending mode in the HOD product. This can be
attributed to a significant change of the water bond angle during the D$^+$
transfer.

%%%%%%%%%%%%%%%%%%%%%%%%%%%%%%%%%%%%%%%%%%%%%%%%%%%%%%%%%%%%%%%%%%%%%%%%%%%%%%

\subsection{Velocity map imaging with crossed beams}

% crossed beam imaging: useful for neutrals
Compared to the conventional crossed-beam experiments with a rotatable
detector, higher angular resolution and a much more rapid data acquisition is
achieved with an imaging spectrometer. Moreover, such a spectrometer detects
products irrespective of their velocity or scattering angle, i.e.\ it
represents a detector with $4\pi$ solid angle of acceptance. Ion imaging has
been combined with neutral reactive scattering starting with a study of the H
+ D$_2$ reaction \cite{kitsopoulos1993:sci} and is being successfully employed
in a number of laboratories for reactive and inelastic crossed-beam scattering
experiments \cite{heck1995:annu,elioff2003:sci,lin2003:sci,zhang2009:sci}. The
current spectrometers are based on the technique of velocity map imaging
\cite{eppink1997:rsi}, which projects ions with the same velocity parallel to
the detector surface onto the same spot on the detector. It thereby avoids
broadening of the product ion images due to the finite size of the reaction
volume.

% crossed beam imaging for ion-molecule reactions
To study ion-molecule reactions with crossed beam imaging, we have constructed
a velocity map imaging spectrometer and a versatile low-energy ion source
\cite{mikosch2006:pccp}. This approach is in contrast with an early
exploration of ion-molecule crossed beam imaging, where an in-situ production
of the reactant ions was used \cite{reichert2002:jcp}, an approach that is not
usable for negative ions. In our experiment, slow ions with between 0.5 and
5\,eV kinetic energy are brought to collision with neutral molecules in a
supersonic molecular beam in the centre of the velocity map imaging electrode
stack. Both reactant beams are pulsed to avoid a heavy gas load in the vacuum
system. Once the two beams have crossed, the electric field of the imaging
spectrometer is rapidly pulsed on and any product ions are projected onto the
position sensitive imaging detector. Ion impact positions, which are
proportional to the velocity components in the scattering plane parallel to
the detector surface, are recorded with a CCD camera. Recently, we improved
the ion imaging electrodes to enhance the velocity resolution and we included
a photomultiplier tube to measure the third, vertical, component of the
product velocity vector \cite{trippel2009:jpb}.

% sn2 experiment
With the ion-molecule crossed beam imaging spectrometer we have studied the
elementary S$_{\rm N}$2-reaction of Cl$^-$ with methyl iodide (CH$_3$I) at
relative scattering energies between 0.4 and 2\,eV \cite{mikosch2008:sci}. The
potential energy curve for this reaction along the reaction coordinate is
closely resembled in Fig.\ \ref{reaction-pot:fig}. As discussed already in
section \ref{interaction:sect} the characteristic deep potential minima,
separated by an intermediate barrier, are found. In the reaction studied here
this barrier lies submerged below the energy of the reactants. Nevertheless it
strongly influences the reactivity in that the reaction occurs at only 10\% of
the Langevin or capture rate.

% data analysis
The imaging data, which is represented by an event list of impact velocity
vectors in the laboratory frame, needs only little data processing. On the one
hand, the transformation to the centre of mass frame has to be performed for
each scattering event, which involves a translation and a rotation of the
image parallel to the relative velocity vector. On the other hand, the loss of
fast moving reaction products needs to be corrected. Product ions with high
laboratory velocities have a chance to leave the spectrometer volume that can
be imaged onto the position-sensitive detector. This effect is referred to as
'density-to-flux correction' in neutral crossed beam scattering experiments,
where it also includes the correction for the spatial and velocity dependence
of the ionisation efficiency. For ion-molecule reactive imaging, the same
correction function is used for all relative collision energies. It only
depends on the magnitude of the laboratory velocity and not on its direction
in the lab frame.

%%% Figure
\begin{figure}[tb]
\center
\includegraphics[width=\columnwidth]{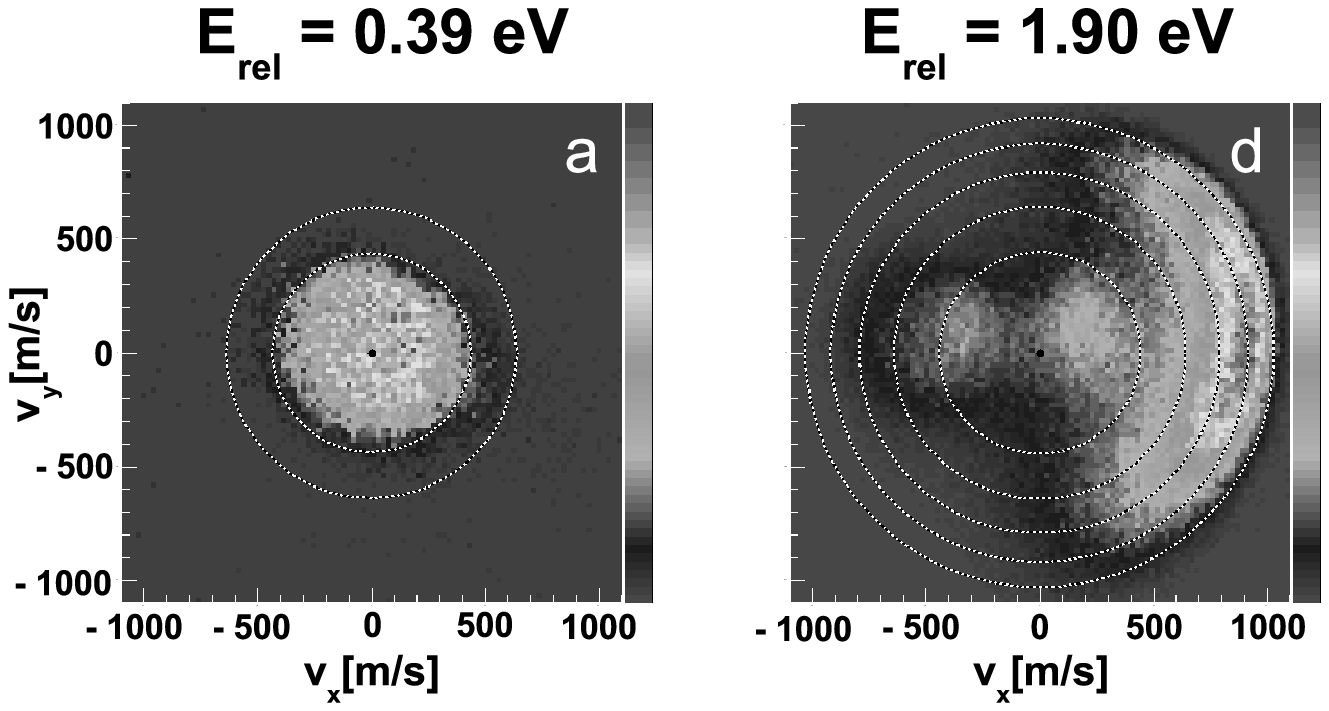}
\caption{\label{sn2-images:fig} Measured differential scattering cross section
  for the reaction of Cl$^-$ + CH$_3$I giving ClCH$_3$ + I$^-$. Shown are two
  images at 0.39 (left panel) and 1.9\,eV (right panel) relative collision
  energy of the I$^-$ velocity vector in the scattering plane obtained by
  velocity map slice imaging. The centre of each image denotes zero velocity
  in the centre of mass frame. The circles represent constant product
  velocities with the largest circle showing the maximum possible product
  velocity based on the known total energy in the reaction system. At 0.39\,eV
  isotropic scattering is observed, indicative of an indirect reaction
  mechanism via a long-lived complex. At 1.9\,eV most of the flux shows direct
  scattering with large product velocities (peak near $v_x \sim +1000$\,m/s,
  but with about 10\% probability, small product velocities both forward and
  backward scattered, are observed. These events are attributed to the
  indirect ``roundabout'' mechanism (Taken from \cite{mikosch2008:sci}).}
\end{figure}

% data
Two measured images of the differential cross section in the centre-of-mass
frame are shown in Fig.\ \ref{sn2-images:fig} for two relative energies
\cite{mikosch2008:sci}. At 0.39\,eV isotropic scattering of the I$^-$ product
ion is observed, indicative of an indirect reaction mechanism via a long-lived
complex.  Here, also much more energy is partitioned into internal excitation
of the neutral CH$_3$Cl product. This indicates an indirect reaction mechanism
with trapping of the collision partners in the minima of the intermolecular
reaction potential. In contrast, at 1.9\,eV the I$^-$ ions scatter
preferentially backward with respect to the direction of the incoming
CH$_3$I. Also their velocity is found very near the maximum possible
velocity. This is explained by a fast and direct reaction mechanism where the
I$^-$ leaves the reaction approximately co-linearly with the incoming Cl$^-$
anion.

% sn2 theoretical description
In order to understand the details of the measured differential cross sections
theoretical calculation are employed. However, the theoretical description of
polyatomic reactions that involve more than four atoms is very difficult. The
present reaction involves six atoms and therefore twelve internal degrees of
freedom. Such a large system can not be calculated quantum mechanically and
one has to resort to significant approximations. These are either quantum
scattering calculations in reduced dimensions (typically four)
\cite{schmatz2004:cpc} or calculations that treat the electronic
structure quantum mechanically but propagate the nuclei classically on the
Born-Oppenheimer surface \cite{hase1994:sci}.

% sn2 theoretical results
Calculated trajectories show that at 1.9\,eV collision energy a direct
reaction mechanism governs the nucleophilic substitution reaction. The Cl$^-$
ions moves into the umbrella of the hydrogen atoms and forms a bond with the
central carbon atom. Roughly co-linearly to this motion the I$^-$ product ion
is moving away after the three hydrogen atoms have inverted to form the
ClCH$_3$ product molecule. This numerical results corresponds directly to the
back-scattering observed in the experiment. The trajectory calculations also
revealed another reaction mechanism that occurs with about 10\%
probability. In this mechanism the CH$_3$I reactant undergoes a single
360$^\circ$ revolution about an axis perpendicular to the C-I bond. Only after
this revolution the substitution occurs. This mechanism, which we named the
``roundabout'' mechanism, is found to go along with a large energy
partitioning into internal degrees of freedom of the neutral product molecule.
This agrees with the observed structures in the measured differential cross
sections at small I$^-$ velocity (see Fig.\ \ref{sn2-images:fig}), which have
therefore been attributed to the roundabout mechanism \cite{mikosch2008:sci}.

%%%%%%%%%%%%%%%%%%%%%%%%%%%%%%%%%%%%%%%%%%%%%%%%%%%%%%%%%%%%%%%%%%%%%%%%%%%%%%
%%%%%%%%%%%%%%%%%%%%%%%%%%%%%%%%%%%%%%%%%%%%%%%%%%%%%%%%%%%%%%%%%%%%%%%%%%%%%%
%%%%%%%%%%%%%%%%%%%%%%%%%%%%%%%%%%%%%%%%%%%%%%%%%%%%%%%%%%%%%%%%%%%%%%%%%%%%%%

\section{Perspectives}

Negative ion reactions have been studied for a long time owing to their
importance in many Earth-bound, planetary or astrophysical plasmas. In the
last two decades research on the detailed dynamics of this class of reactions
has flourished, owing to more and more precise techniques to measure both
integral and differential scattering studies. In this article we have
presented an overview of these experimental approaches and how much they have
been teaching us on the different, and sometimes peculiar, aspects of
anion-molecule reaction dynamics.

We have shown that flow and drift tube measurements are well suited to study
the thermal kinetics of anion-molecule reactions. At low temperature these
studies are complemented and extended using cryogenic ion traps, which also
can measure much smaller rate coefficients than drift tubes due to the long
attainable interaction times. Guided ion beam experiments are ideal for
precise measurements of integral cross sections at well-defined relative
energy over a large dynamic range. The dynamics of the transient reaction
complex can be studied with direct time-resolved spectroscopy and its lifetime
can also be indirectly inferred from ternary collision rates. When it comes to
direct imaging the full reaction dynamics, crossed-beam experiments are very
revealing. In particular the new opportunities of crossed-beam velocity map
ion imaging should be stressed here.

Despite a wealth of theoretical studies, detailed insight into the flow of
energy via the coupling of different vibrational modes during the reaction
continues to be not accessed experimentally.  Of particular interest is the
regime of breakdown of ergodicity at the transition from a complex mediated,
statistical to an impulsive reaction mechanism. It is in this range of
relative collision energies, where crossed beam imaging provides the best
resolution.  Reactive scattering with vibrationally excited reactants - as
successfully used to shed light on chemical reactions of neutrals
\cite{zhang2009:sci} - is not yet explored for anion-molecule reactions. The
role of non-reactive degrees of freedom in a reaction, referred to as the
``spectator modes'', on the integral cross section and the product branching
is a matter of debate that needs answers from experiments. Reactive scattering
with spatially aligned or oriented molecules would directly access the
stereodynamics on the molecular level. Importantly, anion-molecule reactions
have very different kinetics in solution. A bottom-up approach, introducing
the solvent molecule per molecule in scattering experiments with microsolvated
anions, would highlight the role of the environment on the dynamics. Questions
about the role of quantum effects such as Feshbach scattering resonances,
tunnelling, zero-point vibrational motion or decoherence free subspaces can be
dared to be asked now and demand innovative experimental approaches.

We expect that guided ion beam and ion trap studies as well as ion-molecule
crossed-beam imaging will be most helpful to address some of these questions
in the future. The combination of an ion trap as a source for internally cold
molecular ions and clusters with a scattering experiment will allow for
studies of the reactions of complex molecular systems while maintaining good
control over their internal quantum states. Also merged ion and neutral beams
represent a useful approach to study low-energy reactions, as demonstrated
most recently for the reaction of H$^-$ with H \cite{bruhns2010:rsi}, which is
considered to be important for the formation of H$_2$ in the early
universe. Laser-cooling of the reactants may eventually provide even lower
collision temperatures than buffer gas cooling, an approach that has been
demonstrated for cation-neutral reactions \cite{willitsch2008:pccp}; Os$^-$ is
a possible candidate for anion laser cooling
\cite{bilodeau2000:prl,warring2009:prl}. All in all, there is also a lot of
room for further research on anion-molecule reactions, that challenges the
``complexity limit'' and ultimately allows us to understand and control all the
subtleties of the chemical dynamics of multi-atom molecular systems.

%%%%%%%%%%%%%%%%%%%%%%%%%%%%%%%%%%%%%%%%%%%%%%%%%%%%%%%%%%%%%%%%%%%%%%%%%%%%%%

\section{Acknowledgments}

We would like to thank our collaborators in the experiments on anion-molecule
reaction dynamics, in particular Sebastian Trippel, Rico Otto, and Christoph
Eichhorn. We also wish to thank Dan Neumark and Bill Hase for the fruitful
collaborations on reaction dynamics in general and nucleophilic substitution
reactions in particular. Our research is supported by the Elitef{\"o}rderung
der Landesstiftung Baden-W{\"u}rttemberg, the Alexander von Humboldt
Foundation, the Deutsche Forschungsgemeinschaft and the EU Marie Curie Initial
Training Network ICONIC.

%%%%%%%%%%%%%%%%%%%%%%%%%%%%%%%%%%%%%%%%%%%%%%%%%%%%%%%%%%%%%%%%%%%%%%%%%%%%%%
%%%%%%%%%%%%%%%%%%%%%%%%%%%%%%%%%%%%%%%%%%%%%%%%%%%%%%%%%%%%%%%%%%%%%%%%%%%%%%
%%%%%%%%%%%%%%%%%%%%%%%%%%%%%%%%%%%%%%%%%%%%%%%%%%%%%%%%%%%%%%%%%%%%%%%%%%%%%%

\end{document}